\documentclass[12pt]{aa}  
\usepackage{graphicx, xcolor}
\usepackage{txfonts}
\usepackage{float}
\usepackage{dblfloatfix}
\bibpunct{(}{)}{;}{a}{}{,}
\usepackage{natbib,twoopt}
\usepackage{bm}
\usepackage{url}
\usepackage[hidelinks]{hyperref}

\usepackage{array}
\usepackage{booktabs}
\hypersetup{
    colorlinks=true,
    linkcolor=blue,
    filecolor=magenta,
    citecolor=blue,
    urlcolor=blue,
    breaklinks=true
    }

\begin{document}

   \title{Galactic foreground residue biases in cosmic-microwave-background lensing-convergence reconstruction and delensing of B-mode maps}

   \author{Kishan Deka\inst{1} \thanks{Corresponding author:         \href{mailto:kishan.deka@ncbj.gov.pl}{\texttt{kishan.deka@ncbj.gov.pl}}}
         \and
          Pawe\l~Bielewicz\inst{1}
          }

   \institute{National Centre for Nuclear Research,
              Pasteura 7, 02-093, Warsaw, Poland 
             }

   \date{Received 04 November 2025, Accepted 23 February 2026}
   \abstract{

   Diffuse contamination from Galactic foreground emission is one of the main concerns for reconstruction of the cosmic-microwave-background (CMB) lensing potential for next-generation CMB polarisation experiments. Using realistic simulations, we investigated the impact of Galactic foreground residuals from multi-frequency foreground-cleaning methods on CMB lensing reconstruction and the de-lensing of B-mode maps. We also assessed how these residuals affect constraints on the tensor-to-scalar ratio for a CMB-S4–like experiment. We paid special attention to studies of the errors coming from the small angular scale non-Gaussianity of the foreground residuals. We show that component separation is essential for the lensing reconstruction that reduces Galactic emission contribution to the lensing reconstruction errors by one order of magnitude. The residual foreground contribution is dominated by terms coming from Gaussian components of the residual maps. Errors coming from non-Gaussian components are around three orders of magnitude smaller than the Gaussian one, even for recent and the most complex models of the Galactic emission considered in this work. Although the bias in the reconstruction errors due to the Gaussian component of the residuals being small, it is comparable to the cosmic-variance limit for the lensing power spectrum. For this reason, we corrected for this bias in the de-lensing of B-mode maps and constraining the tensor-to-scalar ratio. We also show that for the delensed B-mode maps with a simple quadratic estimator, that is, residuals of the Galactic emission after component separation, errors are two orders of magnitude smaller than uncertainties from leftover of the lensing signal. However, for high-sensitivity CMB experiments and more efficient de-lensing algorithms that remove up to 90\% of the lensing signal, the foreground residuals will become one of the main sources of errors. 
   }

    \titlerunning{Galactic foreground biases in CMB lensing reconstruction and delensing of B-modes}
    \authorrunning{Deka, K. \& Bielewicz, P.} 
    
    \maketitle
%
%-------------------------------------------------------------------
\section{Introduction}
    The cosmic-microwave-background (CMB) photons, which decoupled from the baryon-photon plasma approximately 400 000 years after the Big Bang, encode information about the primordial fluctuations that seeded the large-scale structure of the Universe. In the previous decade, the Planck experiment successfully mapped the temperature anisotropies of the CMB with high precision, reaching sensitivity close to the cosmic-variance limit over a wide range of multipoles \citep{Planck_2020_I,Planck_2020_V}. In the new era of precision cosmology, next-generation experiments such as the Simons Observatory (SO; \citealt{SO_2020}), CMB-S4 \citep{cmbs4_2019}, and LiteBIRD \citep{Litebird_2019} are targeting the CMB polarisation field with unprecedented sensitivity up to a resolution scale of a few arcminutes for the first two experiments and around 20 arcminutes for the latter.  
    
    Inflationary models predict the production of the primordial gravitational waves (GWs) in the early Universe that imprint a large-scale divergence-free pattern --so-called B-modes-- in the CMB polarisation field \citep{Starobinskii_1979, Guth_1982, Linde_1983, Fabbri_1983, Abbott_1984, Polnarev_1985, Kamionkowski_1997, Seljak_1997}. Detecting these tensor B-mode signals is one of the primary science goals of current and upcoming CMB experiments. The detection of such signal would provide indirect evidence of inflation. Because no detection has been made far, only an upper limit on the amplitude of the primordial GWs can be constrained. Conventionally, this amplitude is expressed in terms of the so-called tensor-to-scalar ratio, $r$, which is defined as a ratio of the tensor (primordial GWs) and scalar perturbation power spectrum amplitudes for a characteristic pivot scale. Combined measurements from the BICEP/Keck array, Planck, and WMAP experiments give the current upper bound (at $2\sigma$ level) on the tensor-to-scalar ratio as $r<0.036$ at 0.05 Mpc$^{-1}$ pivot scale \citep{BICEP_2021_XIII}.

    The detection of tensor B-modes is challenged by secondary effects such as extragalactic and Galactic foreground contaminations and weak gravitational lensing of the CMB. The intervening matter distribution deflects the CMB photons along its path, remapping the CMB anisotropies. Thus, weakly lensed CMB anisotropies trace the integrated matter density distribution of the Universe along the line of sight to the last scattering surface. Standard deviation of the deflection angle caused by the lensing effect is $\sim 2$ arcminutes when the lensing potential wells along the path are not correlated \citep{Lewis_2006}. The line-of-sight integrated deflection field can be reconstructed using quadratic estimators (QEs) - a method that is described well in the literature exploits the lensing-induced correlations present in observed CMB maps \citep{Hu_2002, Okamoto_2003, Planck_2014_XVIII, Carron_int_2017, Carron_2022}.
    
    The lensing effect can introduce an additional B-mode polarisation pattern by re-mapping and distorting the primordial rotation-free polarisation patterns -- so-called E-modes \citep{Zaldarriaga_1998, Kesden_2002, Sayre_2020}. For the current upper bound on the tensor-to-scalar ratio and high signal-to-noise observations of polarisation fields, subtracting the lensing effect from B-mode polarisation becomes essential for the detection of the tensor B modes \citep{Seljak_2004,Smith_2012, Simard_2015, Sherwin_2015, Baleato_Lizancos_2022}. The subtraction of the lensing effect from observations, often called de-lensing, can be performed using a template of lensing B modes, which is constructed from an observed E mode and a high signal-to-noise estimate of the deflection field on small and intermediate scales, respectively \citep{Manzotti_2017, Namikawa_2017, BICEP_2021_XII, Planck_2020_VIII}.

    With increasing signal-to-noise ratio of current  and forthcoming CMB polarisation experiments, the polarisation-based lensing of potential power spectrum estimators, which uses the higher order statistics of the CMB polarisation field, has become more important \citep{Namikawa_2022, Belkner_2024}. However, these estimators are sensitive to secondary sources of polarised emission such as radio galaxies and dusty star-forming galaxies, which can bias the lensing reconstruction and therefore the delensed B-mode signal \citep{Ferraro_2018, Schaan_2019, Sailer_2021, Sailer_2023}. A diffused Galactic foreground, and particularly polarised emission from thermal dust and synchrotron emission, pose an additional hindrance to delensing of CMB B-modes. These foregrounds display spatial  variations and strong frequency dependence, and they are inherently non-Gaussian on both large and small scales \citep{Gold_2011, Fauvet_2012, Planck_2015_X, Remazeilles_2015, Planck_2015_int_XXII, Planck_2020_IV, Planck_2020_XI}. The foreground contamination can introduce spurious correlations, which will bias both the lensing reconstruction and the estimation of the tensor-to-scalar ratio. Multi-frequency component-separation techniques can mitigate some of these effects, although residual foreground contamination can still propagate into delensed CMB maps \citep{Fantaye_2012,Beck_2020, Belkner_2024, Hertig_2024, Irene_2025, Bianchini_2025}. 

     In this work, we investigated the impact of Galactic foreground on the reconstruction of CMB-lensing power spectra, the de-lensing of B-mode maps, and the resulting constraints on the tensor-to-scalar ratio in detail. Particular attention was paid to the effects of non-Gaussian small-scale Galactic emissions. We quantified the biases introduced in the lensing reconstruction noise and the estimated lensing power spectra for three foreground models with varying degrees of small-scale non-Gaussianity. In this work, we did not include extragalactic foregrounds, as their impact on CMB-lensing reconstruction and de-lensing has already been widely studied in the literature, and in some cases mitigation methods have already been developed.
     
    Using realistic simulations for a next-generation CMB-S4–like experiment, we explored how increasing foreground complexity affects the constraints on primordial gravity amplitudes. The paper is organized as follows. Section \ref{sec:thoery} presents the theoretical framework for component separation, lensing reconstruction, de-lensing, and the overall analysis methodology. The simulations employed are described in detail in Sect. \ref{sec:simulation}. Section \ref{sec:results} discusses the impact of residual Galactic foreground on de-lensing efficiency and the resulting constraints on the tensor-to-scalar ratio. In Sect. \ref{sec:conculsion}, we conclude our findings, discuss the limitations of our work, and provide future prospects of an extension of our work.
    
%--------------------------------------------------------------------
\section{Theoretical background and methodology}   \label{sec:thoery}
    The CMB photons are deflected by the intervening matter distribution of the Universe along its journey from the last scattering surface to us. Weak lensing breaks the statistical isotropy of primordial CMB signal introducing coupling between different spherical harmonic modes. As a consequence of lensing, the CMB temperature and polarisation fluctuations observed in the direction $\boldsymbol{\hat{n}}$ in the sky are coming from primordial fluctuation fields in the $\boldsymbol{\hat{n}} + \nabla \phi (\bm{\hat{n}})$ direction, where $\nabla \phi (\bm{\hat{n}})$ is a deflection angle. Thus, the lensing causes the re-mapping of the primordial fluctuation fields as given by \citep{Lewis_2006}    \begin{equation}
        \begin{alignedat}{1}
        T(\bm{\hat{n}}) &= \tilde{T}(\bm{\hat{n}} + \nabla \phi (\bm{\hat{n}})) \\
        [Q + iU](\bm{\hat{n}}) &= [\tilde{Q}+i\tilde{U}](\bm{\hat{n}} + \nabla \phi (\bm{\hat{n}})).
        \end{alignedat}
    \end{equation}
    Here, $T$ denotes the lensed CMB temperature anisotropies and $Q$, $U$ are the Stokes parameters of lensed CMB polarisation fields. Meanwhile, $\tilde{T}, \tilde{Q}, \tilde{U}$ represents the unlensed CMB fields. The lensing-potential field, $\phi(\bm{\hat{n}}),$ is defined by \citep{Lewis_2000, Hu_2002, Okamoto_2003}
    \begin{equation}
        \phi(\bm{\hat{n}}) = -2 \int_0^{\chi^*} d\chi \frac{D_A(\chi^* - \chi)}{D_A(\chi)\,D_A(\chi^*)} \Psi(\chi(\bm{\hat{n}}), \eta),
    \end{equation} 
    which is the line-of-sight projection of the Weyl gravitational potential, $\Psi(\chi(\hat{n}))$, integrated along the comoving distance $\chi$, from the observer to the last scattering surface at the comoving distance $\chi^*$ ($\approx 14000$ Mpc). The quantity $\eta$ denotes the conformal time when the photon was at distance of $\chi$ in the direction $\bf{\hat{n}}$. $D_A(\chi)$ is the angular diameter distance that depends on the curvature of the Universe. For a flat Universe, the angular diameter distance is $D_A(\chi) = \chi$.

    The unlensed temperature and polarisation fields and the lensing-potential field are assumed to be Gaussian and statistically isotropic. Hence, the statistical properties can be characterized by using the angular power spectra defined as \citep{Zaldarriaga_1997}
    \begin{align}
        \left< \tilde{X}_{\ell m}^* \tilde{Y}_{\ell' m'}^{} \right>  &= \delta_{\ell  \ell'} \delta_{m m'} \tilde{C}_\ell^{XY} \label{eq:power_spectrum_cmb} ,\\
        \left< \phi_{\ell m}^* \phi_{\ell' m'}^{} \right>  &= \delta_{\ell  \ell'} \delta_{m m'} C_\ell^{\phi\phi},  \label{eq:power_spectrum_phi}
    \end{align}

    \noindent where $\tilde{X}_\ell^m$ and $\phi_\ell^m$ are spherical harmonic coefficients of unlensed $\tilde{X} \in  \{\tilde{T},\tilde{E},\tilde{B}\}$ fields and the lensing-potential field $\phi$, respectively. The brackets $\big<.\big>$ denote the ensemble average over multiple realisations, and $\delta_{ij}$ is the Kronecker delta symbol. The observed angular power spectra of $T,E,$ and $B$ also contain instrumental noise characterised by power spectra deconvolved for a beam:
    \begin{align}
        N_\ell^{TT} & = \sigma_T^2 \exp{\frac{\ell(\ell+1)\theta^2_{\text{fwhm}}}{8\log{2}}} = \frac{\sigma_T^2}{B_\ell^2} \label{eq:tnoise} ,\\
        N_\ell^{XX} & = \sigma_P^2 \exp{\frac{\ell(\ell+1)\theta^2_{\text{fwhm}}}{8\log{2}}} = \frac{\sigma_P^2}{B_\ell^2} \label{eq:pnoise},
    \end{align}
    where $XX \in \{EE,BB\,\}$, $B_\ell$ is the beam-transfer function for a Gaussian profile, $\theta_{\text{fwhm}}$ is the angular resolution of the instrument expressed as the full width at half maximum (FWHM) of the beam, and $\sigma_T$ and $\sigma_P$ are the sensitivities corresponding to the intensity and polarisation detectors, respectively. The specification of noise spectra varies for different wavelength observations.

\subsection{Component separation}    
    Diffuse Galactic emission is one of the main sources of contamination in precise CMB measurements. The polarised Galactic foreground emission consists of two components that includes thermal dust emission and synchrotron radiation. Due to the alignment of dust grains with respect to the Galactic magnetic field (GMF; \citealt{Stein_1966, Roger_1988}), the emission is partially linearly polarised \citep{Planck_2014_XI, Planck_2015_int_XXII, Planck_2020_XI}. The synchrotron emission is generated by relativistic electrons in the interstellar medium that spiral in the GMF. It is strongly polarised and dominates at low microwave frequencies \citep{Haslem_1982, Bennett_2003, Gold_2011,Planck_2016_XXV,Planck_2020_IV,Martire_2022}. 
    
    Multi-frequency observations and accurate component separation methods are essential for cleaning CMB signal from the Galactic foreground. A widely used method known as internal linear combination (ILC) recovers the CMB signal as a linear combination of multi-frequency maps with weights that minimise contributions from components of the maps other than the CMB \citep{Bennett_2003}. The sum of the weights is constrained to be equal to one, which ensures that the clean CMB output map has unit response. This method of component separation is blind in the sense that it does not assume any prior knowledge about the foreground emissions. \citet{Bennett_2003} showed the implementation of this method for the Wilkinson Microwave Anisotropy Probe (WMAP) temperature maps. \citet{Eriksen_2004} extended the analysis to polarisation maps through the use of Lagrange multipliers.
    Similarly to pixel-based ILC, the harmonic ILC (HILC) method uses the spherical-harmonic coefficients of CMB anisotropies in a linear combination of the coefficients of different frequency maps \citep{Tegmark_2003,Kim_2009}. An advantage of the HILC method is that the Harmonic ILC weights depend on angular scale, whereas the pixel-based ILC method provides scale-independent ILC weights.
    \vspace*{0.1cm}
    
\subsection{Quadratic estimator} \label{sec:sec2.2}
    Following the seminal work by \citet{Hu_2002} and its generalisation for the full sky analysis by \citet{Okamoto_2003}, we implemented the quadratic estimator (QE) to reconstruct the lensing-potential field. The lensing of CMB photons breaks the statistical isotropy of CMB fluctuations introducing coupling between spherical harmonic modes. This gives rise to non-zero off-diagonal covariance between two lensed CMB fields, which up to first order in the lensing potential, $\phi,$ is given by 
    \begin{equation}
        \left<X_{\ell m} Y_{\ell'm'}\right>_{CMB} = \sum_{LM} (-1)^M \begin{pmatrix}
            \ell & \ell' & L \\
            m & m' & -M
        \end{pmatrix}\,
        f_{\ell L \ell'}^{XY}\, \phi_{LM}
        \label{eq:lens_correlation}
    .\end{equation}
    The ensemble average $\left< . \right>_{CMB}$ is given over the lensed CMB fields assuming, contrary to the average introduced in Eq. (\ref{eq:power_spectrum_phi}), a fixed lensing-potential field. The lensing response functions, $f^{XY}_{\ell  \ell'L}$, for different pairs of fields are shown in Table (\ref{tab:lens_response}). The coefficients, $F^{(s)}_{\ell L \ell'}$, are defined as the mode-coupling ones due to the lensing \citep{Okamoto_2003}. In Eq. (\ref{eq:lens_correlation}), we use the Wigner-$3j$ symbol representation \citep{Okamoto_2003, Smith_2012}.

    \begin{table}[]
        \centering
        \caption{The lensing response function, $f_{\ell_1 L \ell_2}$, for different $XY$ field pairs.}
        {\renewcommand{\arraystretch}{1.3}
        \begin{tabular}{c c} 
            \hline \hline
            $XY$ & $f^{XY}_{l_1Ll_2}$ \\ \hline
             $TT$ & $\tilde{C}_{\ell_1}^{TT}F^{(0)}_{l_2Ll_1} + \tilde{C}_{\ell_2}^{TT}F^{(0)}_{l_1Ll_2}$ \\
             $EE$ & $\tilde{C}_{\ell_1}^{EE}F^{(2)}_{l_2Ll_1} + \tilde{C}_{\ell_2}^{EE}F^{(2)}_{l_1Ll_2}$ , even \\
             $TE$ & $\tilde{C}_{\ell_1}^{TE}F^{(2)}_{l_2Ll_1} + \tilde{C}_{\ell_2}^{TE}F^{(0)}_{l_1Ll_2}$ , even \\
             $TB$ & $i \tilde{C}_{\ell_1}^{TE}F^{(2)}_{l_2Ll_1}$ , odd \\
             $EB$ & $i \tilde{C}_{\ell_1}^{EE}F^{(2)}_{l_2Ll_1}$ , odd \\
        \end{tabular}
        }
        \tablefoot{We considered the unlensed $\tilde{C}_\ell^{BB} \sim 0$. 'Odd' and 'even' denote whether the function is non-zero when $l_1+l_2+L$ is odd or even, respectively.}
        \label{tab:lens_response}
    \end{table}
       
    An estimate of the lensing-potential field can be obtained by taking a weighted sum of quadratic combination of filtered observed fields as \citep{Hu_2002, Okamoto_2003}
    \begin{equation}
        \hat{\phi}_{LM}^{XY} = \frac{A_L^{XY}}{L(L+1)} \sum_{\ell m}\sum_{\ell^\prime m^\prime} (-1)^M \begin{pmatrix}
             \ell & \ell^\prime & L \\
             m & m^\prime & -M 
        \end{pmatrix}
        \mathcal{W}^{XY}_{\ell L \ell'} ~ \bar{X}_{\ell m} \, \bar{Y}_{\ell'm'.}  \label{eq:lens_qest}
    \end{equation}
    Here, $\bar{X}_{\ell m}$ and $\bar{Y}_{\ell'm'}$ are inverse variance-filtered observed CMB fluctuation fields. The inverse-variance filtering of an observed field, $\hat{X}_{\ell m}$, in harmonic space is given by
    \begin{equation} \label{eq:inv_weight}
        \bar{X}_{\ell m} = \left(C^{XX}_\ell+N^{XX}_\ell\right)^{-1} \hat{X}_{\ell m}, 
    \end{equation}
    under the assumption that the noise covariance matrix is diagonal. In presence of foreground residual, the inverse weight becomes $(C^{XX}_\ell + N_\ell^{XX} + F^{res}_\ell)^{-1}$, including the residual foreground power spectrum, $F_\ell^{res}$.
    
    The weights, $\mathcal{W}^{XY}_{\ell L \ell^\prime}$, in Eq. (\ref{eq:lens_qest}) are obtained by minimizing the variance $\left< |\hat{\phi}_{LM}|^2 \right>$ assuming a Gaussian distribution of CMB anisotropy. The normalisation factor is defined so that the expectation of the estimator is unbiased:
    \begin{equation}
        \left< \hat{\phi}_{LM} \right> = \phi_{LM}    
    .\end{equation}
    The normalisation factor, $A_L^{XY}$, and the minimum variance weights, $\mathcal{W}^{XY}_{\ell \ell^\prime L}$, are well described in \citet{Okamoto_2003}. 
    The angular power spectrum of the reconstructed lensing field is given by
    \begin{equation}
        \hat{C}_L^{\phi\phi,\, XYX'Y'} = \frac{1}{2L+1} \sum_M \hat{\phi}^{XY\,*}_{LM}\hat{\phi}^{X'Y'}_{LM} \quad ,
    \end{equation}
    which, after averaging over ensemble, can be expressed as
    \begin{equation}
        \left< \hat{C}_L^{\phi\phi,\,XYX'Y'} \right> =  C_L^{\phi\phi} + N_L^{XY X^\prime Y^\prime} .
    \end{equation}
    Here, $N_{L}^{XXX'Y'}$ denotes the reconstruction noise power spectra associated with the quadratic estimator. The reconstruction noise depends on the instrumental noise levels and the resolution of the experiment. The noise term contains different bias contributions with different orders of dependency on $C_L^{\phi\phi}$:
    \begin{equation}
        N_L^{XYX'Y'} = N_L^{(0)} + N_L^{(1)} + N_L^{(2)} + \dots
    .\end{equation}
    In the QE approach, the lensing-power spectrum is essentially estimated from the four-point function of the CMB fields. Following Wick's theorem, the contribution from disconnected four-point contractions gives rise to the zeroth-order noise term, $N_L^{(0)}$, which have no dependency on $C_L^{\phi\phi}$. The $N_L^{(0)}$ noise would be present even in the case of purely Gaussian fields without any lensing effect present. The first-order $N_L^{(1)}$ noise is from connected lensed trispectrum contributions due to secondary lensing effects \citep{Kesden_2003}. The $N_L^{(1)}$ bias has first-order dependency on $C_L^{\phi\phi}$ and becomes significant at small angular scales ($L \gtrsim 1000$). Following \citet{Hanson_2011}, we used lensed theoretical total power spectra in the calculation of $A^{XY}_L$ and $\mathcal{W}^{XY}_{\ell L \ell'}$ in Eq.~(\ref{eq:lens_qest}) to mitigate higher-order bias contributions in $\hat{C}_L^{\phi\phi}$.

    \begin{figure}
        \centering
        \includegraphics[width=1.0\linewidth]{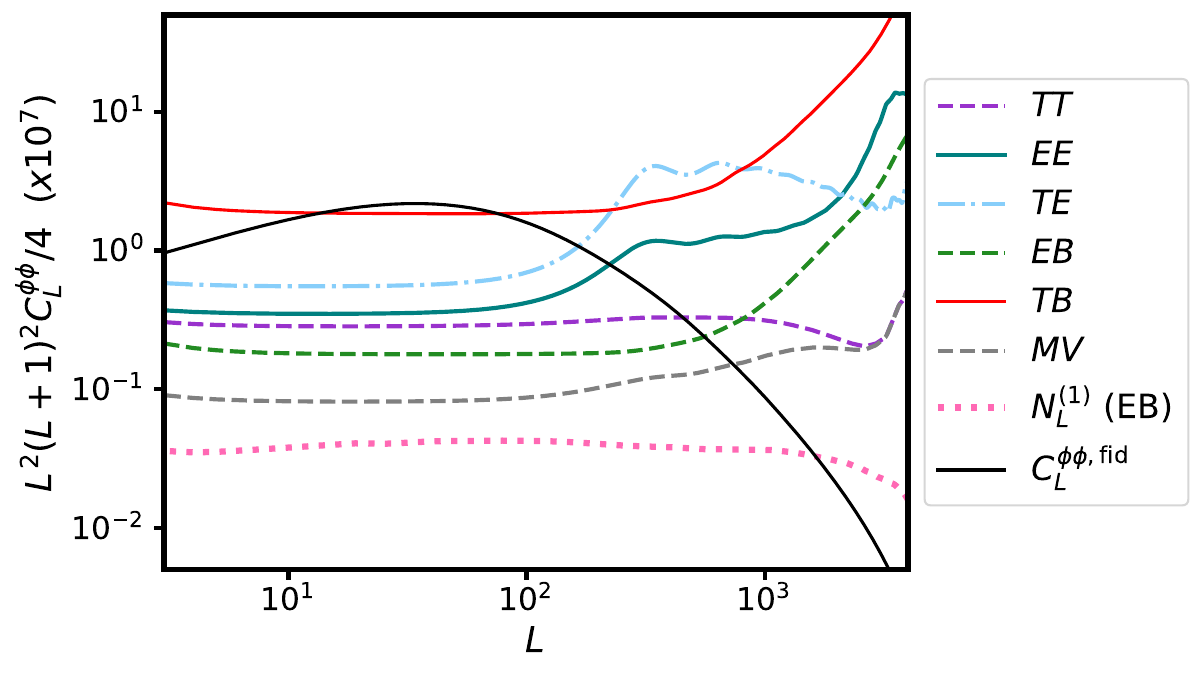}
        \caption{Analytical zeroth-order lensing reconstruction bias, $N_L^{(0)}$, for different estimators given by Eq.~(\ref{eq:nl0_anal_bias}). The instrumental specification was considered for a CMB-S4-like experiment with a beam resolution of 2.5 arcminutes and white-noise temperature level of $\Delta_T$ = 2 $\mu K$ arcmin and polarisation of $\Delta_P$ =  2.8 $\mu K$ arcmin. }
        \label{fig:anal_N0_bias}
    \end{figure}
    
    The disconnected bias contribution term, $N_L^{(0)}$, is expressed analytically as
    \begin{align}
        N^{(0) XYX'Y'}_L \left[ C_\ell \right]
        & = \frac{A_L^{XY*} A_L^{X'Y'}}{2L+1}
       \sum_{\ell \ell^\prime} \mathcal{W}^{XY *}_{\ell L \ell^\prime}
       \left[ \mathcal{W}^{X'Y'}_{\ell L \ell^\prime} C^{XX'}_\ell C^{YY'}_\ell + \right. \nonumber \\ 
        & \quad \left. (-1)^{\ell+\ell^\prime+L} \mathcal{W}^{X'Y'}_{\ell L \ell^\prime} C^{XY'}_\ell C^{X'Y}_\ell \right],
       \label{eq:nl0_anal_bias}
    \end{align}  
    where $C_\ell = \left< \hat{C}_\ell \right>$ is the expectation of estimated total power spectra with noise and foreground contributions. 
    
    To simplify analysis hereafter, we considered identical pairs of fields in both legs of the lensing-power-spectrum estimator; that is,~we assumed that $X=X'$ and $Y=Y'$. Then, the $N_L^{(0)}$ noise follows:
    \begin{equation}
        N_L^{(0) XY}[C_\ell] = A_L^{XY}.
    \end{equation}
    In Fig. \ref{fig:anal_N0_bias}, we show the $N_L^{(0)}$ bias contribution for different QEs for CMB-S4-like experiments. Following \cite{Okamoto_2003}, we also show a minimum-variance (MV) estimator combined from all different estimators that has the highest signal-to-noise ratio. Nevertheless, in our analysis we use only polarisation based EB estimator because it is the least sensitive to the mean-field effect arising from survey inhomogeneities.

    \subsection{Mean-field effect}
\begin{figure*}
    \centering
    \includegraphics[width=0.8\textwidth]{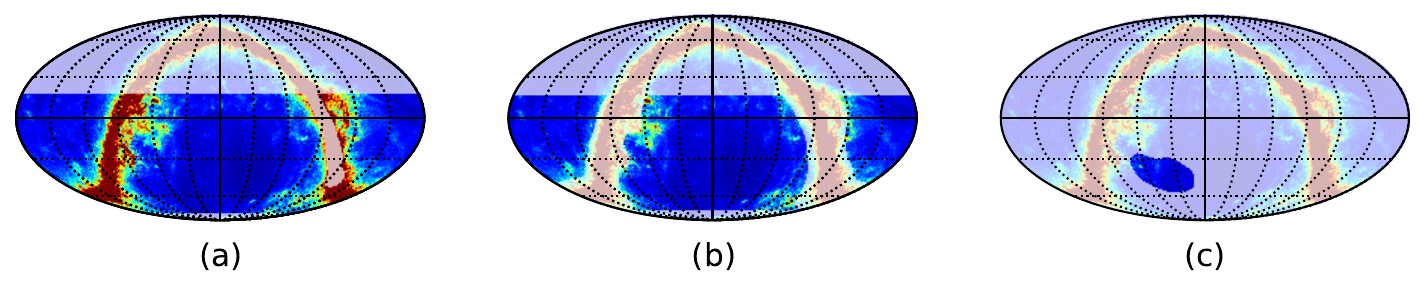}
    \caption{Masks used in our analysis to mimic CMB-S4 sky coverage, shown in Ecliptic coordinates with Galactic emission at 145~GHz in the background.
    (a) The left panel shows the mask used for component separation for the wide sky survey corresponding to the LAT configuration ($f_{\rm sky}=0.64$).
    (b) The middle panel shows the mask used for lensing reconstruction for the wide sky survey ($f_{\rm sky}=0.37$).
    (c) The right panel shows the sky area considered for the deep survey corresponding to the SAT configuration, which was used for the delensing study ($f_{\rm sky}=0.025$).
    }
    \label{fig:skycoverage}
    \end{figure*}

    The reconstructed lensing-potential field using QE is sensitive not only to the anisotropy generated by the lensing effect, but also to other sources of statistical anisotropy. One of the most significant is anisotropy related to the incompleteness of full sky coverage for the ground-based observations and the removal of the parts of the sky most contaminated by Galactic foreground \citep{Planck_2013_XVII, Hanson_2009, Benoit_2013}. For the CMB-S4-like survey, only the southern half of the sky will be observed. The sky coverage is shown in Fig.~\ref{fig:skycoverage}. Besides anisotropy from masking part of the sky, there are also anisotropies induced by inhomogeneous instrumental noise \citep{Hanson_Rocha_2010} and asymmetric beams \citep{Hanson_Lewis_2010}. The latter is the least significant, so we neglected it in our analysis. The remaining anisotropies introduce systematic bias in the reconstructed lensing field, that is, the so-called mean field. We estimated the effect of masking and inhomogeneous noise in the mean-field signal, $\hat{\phi}^{MF}_{LM}$, by taking an average over different realisations of the lensing field reconstructed using the QE, $\left< \hat{\phi}^{QE}_{LM} \right>$. The ensemble average of a random Gaussian lensing field, $\hat\phi_{LM}$, should go to zero, but the presence of anisotropies in the CMB fields used in the estimator leads to non-zero mean-field residuals. To obtain unbiased estimates of the lensing field, we subtracted the mean-field from the quadratic estimate of the lensing potential \citep{Planck_2014_XVIII}:
    \begin{align}
        \hat{\phi}^{MF}_{LM} & = \left< \hat{\phi}^{QE}_{LM} \right> ,\\
        \hat{\phi}_{LM} & = \hat{\phi}^{QE}_{LM} -  \hat{\phi}^{MF}_{LM.}
    \end{align} 

    \subsection{Foreground effects} \label{sec:foreground}
    The presence of Galactic foreground affects lensing reconstruction in two ways. Firstly, it modifies the reconstruction noise power spectrum associated with the QE due to additional power of the foreground signal. In particular, assuming Gaussianity of the foreground, the reconstruction noise defined in Eq. (\ref{eq:nl0_anal_bias}) becomes a function of the total power spectra that includes the foreground angular power spectrum, $F_\ell$, given as \citep{Beck_2020}
    \begin{equation}
        N^{(0),~XY X^\prime Y^\prime}_L \left[ \hat{C}_\ell \right] \rightarrow N^{(0),~XY X^\prime Y^\prime}_L \left[ \hat{C}_\ell + F_\ell \right] \label{eq:n0_anal_foreground}
    .\end{equation}
     On the other hand, non-isotropic distribution of the foreground emission can contribute to the lensing-potential estimation biasing the estimator. The power spectrum of this term, $F_L^{syst}$, is also a systematic bias for the lensing-potential power spectrum \citep{Beck_2020}. As it depends on the connected four-point correlation function of the foreground emission, the bias is very sensitive to the level of non-Gaussianity of the foreground. The $F_L^{syst}$ term can be computed by implementing the quadratic estimator on the Galactic-foreground emission maps. We also considered systematic bias arising from residual foreground present in the HILC cleaned CMB maps and denoted it as $F_L^{syst,~res}$. Similarly to the $F_L^{syst}$ term, $F_L^{syst,~res}$ was estimated from the residual foreground maps.

    \subsection{B-mode template de-lensing}
        In the Taylor expansion of the lensed polarisation fields, the lensed B modes at first order in the lensing potential field become
        \begin{equation}
            B^{len}_{\ell m} = \sum_{l_1 m_1 L M} g_{\ell l_1 l_2}^{EB} \begin{pmatrix}
                \ell & l_1 & L \\
                m & m_1 & M
            \end{pmatrix} E^*_{l_1 m_1} \phi^*_{L M}  \label{eq:lensing_Bmode}
        ,\end{equation}
        which corresponds to convolution between unlensed E modes, $E_{\ell m}$, and lensing potential, $\phi$. The coupling coefficients, $g^{EB}_{\ell l_1 L}$, are defined as \citep{Smith_2012}
        \begin{equation}
            g^{EB}_{\ell l_1 L} = \frac{F^{(-2)}_{\ell l_1 L} - F^{(2)}_{\ell l_1 L}}{2i} .
        \end{equation}
        The mode coupling due to the lensing is captured by $F^{(\pm 2)}_{\ell \ell' L}$ \citep{Smith_2012, Okamoto_2003}. The first-order approximation is sufficient for de-lensing studies at large angular scales relevant for the constraints on the tensor-to-scalar ratio \citep{Smith_2012, Challinor_2005}.
        
        To de-lens observed B modes, a template of lensed B modes can be obtained using the Wiener-filtered observed E modes, $E_{\ell m}^{obs}$, and the Wiener-filtered reconstructed lensing potential, $\hat{\phi}$. Thus, the lensed B-mode template at gradient order is given by \citep{Smith_2012}        \begin{equation}
        \begin{split}
            B^{temp}_{\ell m} = \sum_{\ell' m' L M} g_{\ell \ell' L}^{EB} \begin{pmatrix}
                \ell & \ell' & L \\
                m & m' & M
            \end{pmatrix} & \left( \frac{C^{EE}_{\ell'}}{C^{EE}_{\ell'} + N^{EE}_{\ell'}} \right) E^{obs~*}_{\ell' m'} \\
            & \times \left( \frac{C^{\phi \phi}_{L}}{C^{\phi \phi}_{L} + N^{\phi \phi}_{L}} \right) \hat{\phi}^*_{LM}~.
        \end{split}  \label{eq:lensing_Btemp}
        \end{equation}
        In the presence of foreground residuals, we used residual foreground and noise power spectra in the Wiener filtering of E modes to suppress the contribution of highly contaminated modes.
        Then, the de-lensed B-mode map is given by,
        \begin{equation}
            B^{del}_{\ell m} = B^{obs}_{\ell m} - B^{temp}_{\ell m} \label{eq:delens_Bmode}
        .\end{equation}
        \citet{Baleato_Lizancos_2021} showed that using lensed E modes in Eq. (\ref{eq:lensing_Btemp}) leads to a cancellation of higher order terms in the de-lensed power spectra, thereby reducing the residual lensing floor. Other widely used de-lensing methods that construct B-mode templates from non-perturbative remapping of observed E-modes have been shown to leave one-order-higher residual lensing compared to gradient-order templates \citep{Baleato_Lizancos_2021}. Anti-lensing approaches, in which the observed E modes are first de-lensed by inverse lensing methods and then used to generate a non-perturbatively re-mapped lensed B-mode template \citep{ACT_2020}, achieve similar de-lensing efficiency to gradient-order templates, but they are computationally more expensive. The gradient-order template method adopted in this paper is therefore well suited for forecasting studies due to its analytical simplicity and computational efficiency.
        
        The angular power spectra for de-lensed B-modes can be written as
        \begin{equation}
            C_\ell^{del} = C_\ell^{tens} + C_\ell^{res} + N_\ell^{del} + N_\ell^{res} + F_\ell^{res}  \label{eq:cl_delen}
        ,\end{equation}
        where $C_\ell^{tens}$ is the primordial tensor component power spectrum, $C_\ell^{res}$ is the lensing residual spectra, $N_\ell^{del}$ is the de-lensing bias, $N_\ell^{res}$ is the residual instrumental noise spectra, and $F_\ell^{res}$ is the residual foreground spectra after component separation. The de-lensing bias includes contributions from six-point and higher order correlation functions of CMB fields present in the convolution of observed E-mode and internally reconstructed $\hat{\phi}$ fields using pairs of E- and B-mode maps \citep{Baleato_2021}.
        
        For our analysis, we estimated the residual lensing spectra, $C_\ell^{res}$, from Monte Carlo (MC) simulations of CMB maps by taking the difference between lensing only the B-mode map and lensing B-mode templates over a set of 400 simulations. A power spectrum of the difference also contains the de-lensing-bias contribution, $N_\ell^{del}$, by definition, so this term in Eq. (\ref{eq:cl_delen}) can be omitted. The residual instrumental noise and foreground power spectra were computed by taking an average over a set of 100 simulations for each of the three foreground models described in Sect. \ref{sec:galactic_foreground}. All the power spectra used in our analysis were computed on a cut sky corresponding to the SAT configuration (sky fraction, $f_{sky}=0.025$) using the \texttt{Namaster}\footnote{\url{https://namaster.readthedocs.io/en/latest/}} package, which implements the MASTER algorithm \citep{Hivon_2002, Alonso_2025}.

%-----------------------------------------------------------------

    \subsection{Likelihood analysis} \label{sec:likelihood}
    We used the power spectrum of cleaned and de-lensed B-mode map to constrain the tensor-to-scalar ratio, $r$. In our analysis, we adopted the Hamimeche–Lewis (H-L) likelihood approximation, which closely reproduces the exact full-sky likelihood \citep{Hamimeche_2008}. Unlike a naive Gaussian approximation, the H-L likelihood captures the non-Gaussian statistics introduced by cut-sky analyses and delivers improved accuracy. For a single B-mode polarisation field, the log-likelihood function is written as
    \begin{equation} 
        -2 \ln\mathcal{L}\left(r|\hat{C}_\ell\right) = \sum_{ll^\prime} \left[ g(\hat{C}_\ell/C_\ell)\, C_{fl} \right]
        \left[\textbf{M}_f^{-1}\right]_{ll^\prime} \left[ g(\hat{C}_{\ell^\prime}/C_{\ell^\prime})\, C_{f{\ell^\prime}} \right],
        \label{eq:likelihood}
    \end{equation}
    where
    $$
         g(x) = \text{sign}(x-1) \sqrt{2(x-\ln(x)-1)}~.
    $$
    $[\textbf{M}_f]$ is the covariance matrix estimated from a set of 400 simulated CMB realisations that include a purely Gaussian Galactic foreground. These Gaussian-foreground simulation sets are described in details in Sect. \ref{sec:galactic_foreground}. The covariance matrix computed from a limited number of simulations can be biased \citep{Hartlap_2007}. In our analysis, we used the Hartlap factor to obtain an unbiased estimation of the inverse covariance matrix given as
    \begin{equation}
        \textbf{M}_f^{-1} = \frac{N-d-2}{N-1} \textbf{M}^{-1}_{*}
    ,\end{equation}
    where $N$ is the number of simulations and $d$ is the dimension of the covariance matrix estimator. Here, $\textbf{M}_{*}$ denotes the biased covariance matrix estimator given as
    \begin{equation}
        [\textbf{M}_{*}]_{\ell{\ell^\prime}} = \left< \left(\hat{C}_\ell - C_{fl}\right) \left(\hat{C}_{\ell^\prime} - C_{f{\ell^\prime}}\right) \right> ~.  \label{eq:cov_mat}
    \end{equation}

    The model angular power spectrum is given by 
    \begin{equation}
        C_\ell = r\,C_\ell^{tens,r=1} + C_\ell^{res} + N^{res}_\ell + F^{res}_\ell,  \label{eq:model_cl}
    \end{equation}
    where $C_\ell^{tens,r=1}$ is the theoretical power spectrum of primordial GWs with $r=1$ and spectral index $n_t=0$. We define a fiducial model $C_{fl} = C_\ell$ by setting $r=3\times 10^{-3}$, which is the same as our input value of $r$ for the simulated observations. We considered the covariance matrix to be effectively model independent, as it exhibits no significant variation for changes in $r$ around the fiducial value, and the primordial tensor component contributes negligibly to the uncertainty in the likelihood analysis compared to other components, as shown in Fig.~\ref{fig:sigma_r}.
    
\section{Simulations}   \label{sec:simulation}

    \begin{table}[]
        \large
        \centering
        \caption{Instrumental specifications for polarisation measurements for the CMB-S4--like experiment.}
        \resizebox{0.48\textwidth}{!}{
        \begin{tabular}{cccccccc}
           \hline \hline \\[-0.3cm]
           frequency (GHz) & 20  & 30 &  40 & 95 & 145 & 220 & 270 \\ \hline 
           $\sigma_P$  ($\mu K-arcminute$) & 48.1 & 16.2 & 9. & 1.5 & 1.6 & 5. & 12. \\  \hline
           FWHM ($arcminutes$) & 11. & 7.4 & 5.1 & 2.2 & 1.4 & 1.1 & 0.9  \\ \hline
        \end{tabular}
        } \vspace*{0.2cm}
        \label{tab:noiseprop}
    \end{table}

    To simulate realistic sky observations, we considered lensed CMB maps, instrumental noise, and foreground emission. We calculated theoretical unlensed CMB spectra and lensing potential power spectra with the publicly available CAMB\footnote{\url{https://camb.readthedocs.io/en/latest}} package \citep{Lewis_2000}. We used a standard $\Lambda$-cold dark matter ($\Lambda$-CDM) cosmology with the cosmological parameters from the Planck 2018 release \citep{Planck_2018_VI} : $\Omega_c h^2=0.120, \Omega_b h^2=0.022, n_s=0.965,\tau=0.054,H_0=67.4$. We considered sets of simulations for two tensor-to-scalar ratios: $r=0$ and $r=3\times 10^{-3}$. Using the \texttt{healpy} package\footnote{\url{https://healpix.sourceforge.io/}} \citep{Gorski_2005}, we generated full sky maps of the Stokes parameters $Q$ and $U$ from the obtained theoretical CMB power
spectra and Gaussian random realisations of lensing-potential field from the lensing power spectrum. The CMB maps were then lensed using publicly available \texttt{lenspyx}\footnote{\url{https://github.com/carronj/lenspyx}} package  that provides a curved-sky lensing algorithm for the full sky \citep{Reinecke_2023}. We simulated all maps at a resolution parameter of $N_{\mathrm{side}} = 2048$. The lensed full-sky Stokes parameter maps were converted to full-sky E- and B-mode maps and were then convolved with Gaussian beams corresponding to the seven frequency bands of a CMB-S4--like reference experiment, as listed in Table~\ref{tab:noiseprop}. Subsequently, we added beam-smoothed, simulated E- and B-mode thermal dust and synchrotron polarisation maps to the respective frequency channels. In the Monte Carlo simulations, we kept the Galactic-foreground emission templates fixed and identical across all realisations. We provide the details of the foreground emission maps and their frequency scaling in subsequent section. We then added Gaussian E- and B-mode noise maps and then applied the appropriate survey masks for different telescope configurations.

 \subsection{Galactic foreground} \label{sec:galactic_foreground}
     Linearly polarised thermal dust emission dominates the CMB signal at frequencies higher than $90~GHz$, where upcoming CMB surveys will operate, while the synchrotron emission dominates the polarisation signal at lower frequencies.

    \begin{figure}
        \centering
        \includegraphics[width=1.0\linewidth]{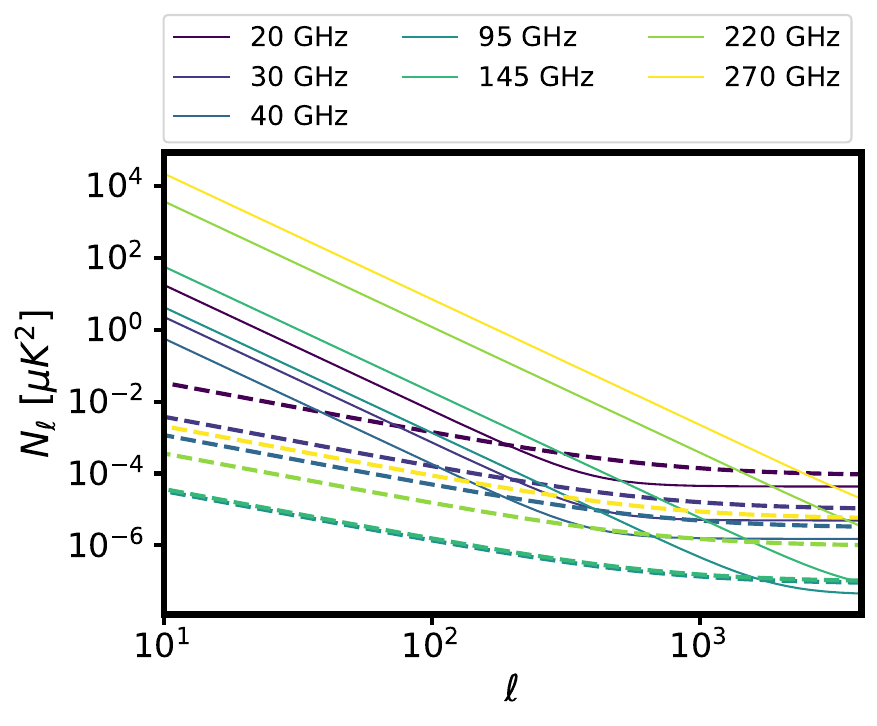}
        \caption{Beam-de-convolved noise power spectra for temperature (solid) and polarisation (dashed) in LAT configuration.}
        \label{fig:cmbs4_noise}
    \end{figure}
        
     The intensity of the thermal dust emission follows the scaling law of a modified-black-body (MBB) emission \citep{Planck_2015_int_XXII}:
     
     \begin{align}
         I^{dust}_\nu & = I^{dust}_{\nu_0} \left(\frac{\nu}{\nu_{0}}\right)^{\beta_{d}} \frac{B_{\nu}(T_d)}{B_{\nu_0}(T_d)}.  \label{eq:dust_scaling}
     \end{align}
     Here, $I^{dust}_{\nu_0}$ is a template of dust emission at a reference frequency, $\nu_0$. The spectral index, $\beta_d$, and dust temperature, $T_d$, have typical values of 1.5 and 20~K, respectively, for both intensity and polarisation, but a spatial variation of the order of 10\% in high Galactic latitude is observed \citep{Planck_2020_XI, Planck_2015_X}. The Stokes parameter $Q$ and $U$ maps also follow the same MBB law with a frequency-dependent polarisation fraction that determines the level of polarisation degree. The polarisation fraction also shows spatial variations and is taken into account in the modelling on small scales using a polarisation-fraction tensor formalism \citep{Borrill_2025}.

     Synchrotron emission arises from relativistic electrons accelerated along the Galactic magnetic field. Synchrotron emission follows a power law \citep{Gold_2011}:
     \begin{equation}
         I^{sync}_{\nu} = I^{sync}_{\nu_0} \left(\frac{\nu}{\nu_0}\right)^{\beta_s}   \label{eq:sync_scaling}
     .\end{equation}
     Similar scaling is applicable to $Q$ and $U$ polarisation fields. Here, the power-law index, $\beta_s$, varies across the sky with values ranging from $\beta_s=-3.0$ near the Galactic plane to $\beta_s=-3.3$ off the plane.
     
     We considered three different pairs of models of Galactic emission, each with different levels of complexity. We used the \texttt{PySM3}\footnote{\url{https://pysm3.readthedocs.io}} package to implement the scaling law for the foreground maps \citep{Thorne_2017, Zonca_2021, Borrill_2025}. The foreground models are listed and described in detail below.

     \paragraph{d1s1} This is the base model where the spectral indices used in the frequency scaling of dust and synchrotron emission have spatial variations. The reference dust templates used are the 545 GHz intensity maps and 353 GHz polarisation maps from the Planck 2015 results \citep{Planck_2015_X}. The dust template maps were smoothed to 2.6 degrees. The synchrotron emission templates for temperature and polarisation were taken from the Haslem maps at 408 MHz \citep{Remazeilles_2015} scaled to 23 GHz and the WMAP nine-year data at 23 GHz \citep{Bennett_2013}. Both templates were smoothed to five-degree resolution. Small scales with Gaussian distributions were then added to both dust and synchrotron maps by fitting a power law to the template maps.

     \paragraph{d10s5}The thermal dust model \texttt{d10} and synchrotron model \texttt{s5} are of medium complexity, with amplitude-modulated small scales chosen to imitate non-Gaussian behaviour. The thermal dust-emission template was taken from the component-separated, thermal dust-emission maps using the Generalised Needlet ILC (GNILC) method from the Planck survey \citep{Planck_2020_IV}. These template maps were obtained at 353 GHz with a resolution of $21.8^\prime$ in terms of temperature and a variable resolution of $21.8^\prime-80^\prime$ for polarisation maps. The \texttt{s5} model is similar to \texttt{s1}, except for the re-scaling of the spectral index map based on the S-band polarisation All Sky Survey (S-PASS; \citealt{Kamionkowski_1997}). Small scales were added to both emission models following the Logarithm of the polarisation-fraction tensor (\texttt{logpoltens}) formalism, which is well described in the \texttt{PySM} documentation \citep{Borrill_2025}. The small scales added have a modulation adopted according to the intensity maps, but we excluded the parts of the sky with the strongest Galactic emission to avoid excessive power in the high-intensity regions of the Galactic plane. The \texttt{logpoltens} formalism and the intensity modulation introduce some non-Gaussianity to the added small angular scales, which is important for our analysis. Additionally, the spectral indices of dust and synchrotron emission and dust temperature maps also contain random fluctuations added to small scales.

     \paragraph{d12s7} The \texttt{d12} model is a three-dimensional line-of-sight-integrated model consisting of six layers, with each layer having a different realisation of dust emission at $\nu_0 = 353$ GHz, different dust temperature maps, and spatially varying spectral indices for frequency scaling \citep{Mart_nez_Solaeche_2018}. On large scales, the model matches observed dust emission, and on smaller scales it is extrapolated from polarised dust-emission power spectra. The small-scale realisations for each layer were modulated on a large-scale polarised intensity level to obtain non-Gaussian statistics. The synchrotron model \texttt{s7} is a more complex version of the \texttt{s5} model, with the introduction of a curvature term to the spectral index in Eq. (\ref{eq:sync_scaling}):     \begin{equation}
         I^{sync}_{\nu} = I^{sync}_{\nu_0} \left(\frac{\nu}{\nu_0}\right)^{\beta_s + C~ln(\nu/\nu_0)}   \label{eq:sync_scaling_curv}
     .\end{equation}
     The curvature term, $C$, was defined for the reference frequency $\nu_0=23~{\rm GHz,}$ which is based on the measurement from the balloon-borne experiment of the Absolute Radiometer for Cosmology, Astrophysics, and Diffuse Emission (ARCADE-2; \citealt{Fixsen_2011, Kogut_2012}).
     The dust model in d12s7 is of the highest complexity, as all individual layers contain intensity-modulated small scales, deviating from Gaussian statistics. The frequency dependency also contributes to non-Gaussianity due to different realisations of spatially varying spectral indices in each layer. The synchrotron emission in d12s7 contains an additional curvature dependence in frequency scaling, which increases its complexity compared to the simple power law in the d1s1 and d10s5 cases.
     \bigskip
    
     To estimate the covariance matrix used in the likelihood analysis (Eq.~(\ref{eq:cov_mat})), in the case of each considered foreground model we used 400 simulations of purely Gaussian foreground maps. We fitted a power law to the power spectra of thermal dust and synchrotron emission at each frequency, and then used these fitted spectra to produce random Gaussian realisations of foreground maps. These maps were added to the CMB realisations along with their corresponding noise realisations, which were then processed through the HILC and de-lensing pipelines to obtain the de-lensed power spectra. Finally, using Eq. (\ref{eq:cov_mat}), we computed the covariance matrices for each model, capturing the off-diagonal contributions arising from mask-induced anisotropies. Additionally, we generated one separate set of simulations with CMB-S4--like noise properties that are free of foreground contamination, but processed in a consistent way through the HILC pipeline; we used these to validate our pipeline. 
    
 \subsection{Sky coverage and instrumental noise} \label{sec:coverage_noise}

    \begin{table}[]
        \centering
        \label{tab:atmospheric_noise}
        \caption{Low-$\ell$ noise model parameters for CMB-S4--like experiment.}
        \begin{tabular}{*5c}
        \hline \hline
        frequency &  \multicolumn{2}{c}{LAT} & \multicolumn{2}{c}{SAT} \\
        \midrule
        $GHz$   & $\ell_{knee}$   & $\alpha$    & $\ell_{knee}$   & $\alpha$  \\ \hline
        20   &  700 & -1.4   & 150  & -2.7 \\
        30   &  700 & -1.4   & 150  & -2.7 \\
        40   &  700  &  -1.4   & 150  & -2.7 \\
        95   &  700 &  -1.4   & 150  & -2.7 \\
        145   &  700 &  -1.4   & 200  & -2.2 \\
        220   &  700  &  -1.4   & 200  & -2.2 \\
        270   &  700  &  -1.4   & 200  & -2.2 \\
        \hline
        \end{tabular}
        \tablefoot{
            The values of $l_{knee}$ and $\alpha$ parameters in Eq. \eqref{eq:lowL_noise} for only polarisation
in both the LAT and SAT configurations.}
    \end{table}

    \begin{figure*}
         \centering
        \includegraphics[width=0.9\textwidth]{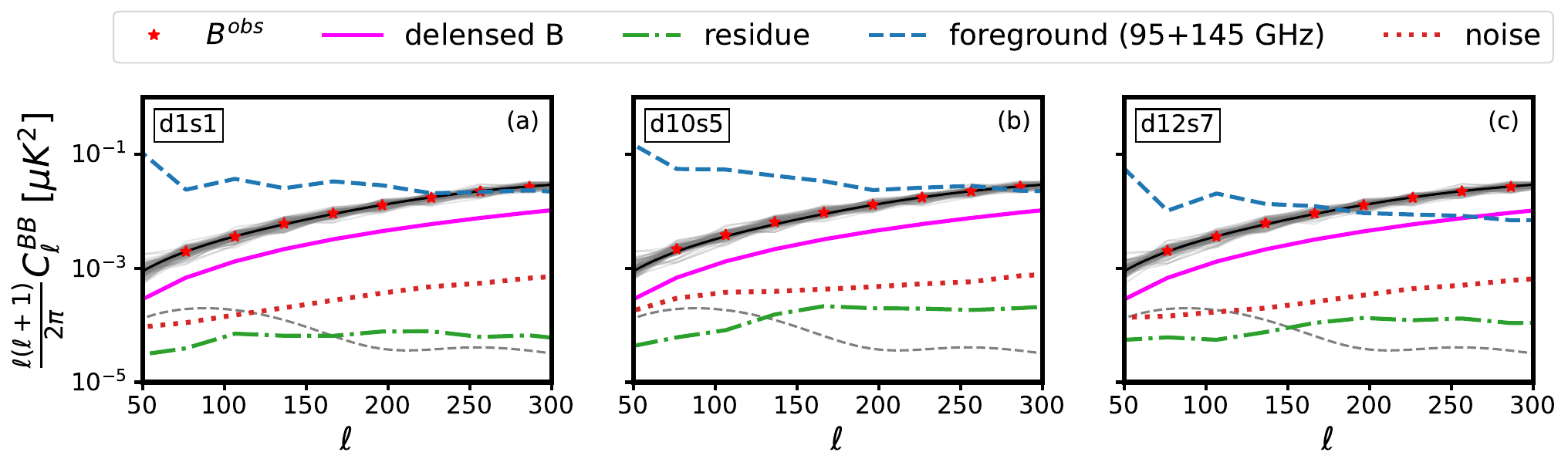}
        \caption{Angular power spectra for CMB B-mode-cleaned maps for the SAT configuration and sky coverage. The solid grey zigzag lines are BB power spectra of 100 simulations, and the average is shown as red stars. The theoretical input power spectra are shown as solid black lines. The residual HILC noise and foreground levels are shown as dotted and dashed lines, respectively. The solid magenta line shows the average de-lensed B-mode spectra over 100 simulations. The dashed blue line shows the average power spectra of Galactic-foreground B modes at the frequency channels 95 and 145 GHz. The dashed grey line shows the tensor B-mode level for $r=0.003$.}
        \label{fig:hilc_sat_ps}
    \end{figure*}

    In our simulated maps, we considered observation of the sky area corresponding to the CMB-S4 experiment. It covers around 60\% of the sky and overlaps with footprints of large-scale structure surveys such as the \textit{Vera Rubin }Observatory (VRO) Legacy Survey of Space and Time (LSST; \citealt{lsst_2009}) and the Dark Energy Spectroscopic Instrument (DESI; \citealt{desi_2016}). Significant overlap is especially important for future cross-correlation studies between the CMB lensing potential and the large-scale structure tracers. We considered the two sky cuts shown in Fig.~\ref{fig:skycoverage} for two separate setups for telescopes used for observations: small-aperture telescopes (SATs) and large-aperture telescopes (LATs). The SATs are considered to perform ultra-deep observations targeting large-scale B-mode signals where primordial B modes are expected to have the highest power. Meanwhile, the LATs observe the small-scale polarisation field in a wide field, which is important to lensing science. The weak lensing information from LATs is relevant to de-lensing analyses and large-scale structure studies.

    The HILC method was implemented on the cut-sky for both LAT and SAT configurations. We applied the corresponding masks to the scalar $E$- and $B$-mode fields directly to mitigate the $E-B$ mixing problem arising from the cut-sky transformation of the Stokes parameters $Q$ and $U$ \citep{Lewis_2001, Bunn_2003}. In actual observations, one has to be careful regarding the $E-B$ mixing problem when working with cut-sky data. For the LAT configuration, we used the mask shown in the left panel of Fig.~\ref{fig:skycoverage}, which retains 65\% of the sky to exclude the highly contaminated Galactic-plane region. In the HILC implementation, we used cut-sky harmonic coefficients, which led to higher residual contamination around the mask edges due to the strong signal gradient at the boundary, which is commonly referred to as the ringing effect \citep{Grain_2009}. We found that this spurious HILC residue near the mask edge introduces additional biases in the lensing reconstruction. To mitigate this, we employed an extended mask by removing a narrow band around the edges of the component-separation mask. We used this final mask in our lensing reconstruction, which covers 37\% of the sky, as shown the middle panel of Fig.~\ref{fig:skycoverage}. For the SAT configuration, we considered a 3\% sky-coverage region located in a low-Galactic-contamination area. To further suppress residual Galactic contamination, we adopted a narrower mask with 2.5\% sky coverage excluding a band around the edges. This mask, shown in the right panel of Fig.~\ref{fig:skycoverage}, was used for the de-lensing and for the likelihood analysis.
    
    The instrumental noise contributions are assumed to be Gaussian. The instrumental properties for different frequencies are listed in Table \ref{tab:noiseprop}. The temperature and polarisation noise power spectra follow Eq. (\ref{eq:tnoise}) and Eq. (\ref{eq:pnoise}), with an additional modification at low multipoles due to atmospheric noise given as
    \begin{equation}
        N_\ell^{X} = \frac{\sigma_{X}^2}{B_\ell^2}\left( 1+\left( \frac{\ell}{\ell_{knee}} \right)^\alpha \right) \quad , \quad X \in \{T,E,B\} 
        \label{eq:lowL_noise}
    .\end{equation}  
    Here, $B_\ell$ is the beam transfer function, and $\sigma_T$ and $\sigma_{E,B} = \sqrt{2}\,\sigma_T$ are the temperature and polarisation noise sensitivity, respectively. We took $\ell_{knee}$ and $\alpha$ parameter values from the CMB-S4 \texttt{DRAFT}\footnote{\url{https://github.com/sriniraghunathan/DRAFT}} tool. The parameter, $\ell_{knee}$, and power-law index, $\alpha$, for SATs and LATs in our polarisation measurements are listed in Table \ref{tab:atmospheric_noise}. In Fig. \ref{fig:cmbs4_noise}, we show the beam-de-convolved noise spectra for different frequencies corresponding to LAT specifications. We used the  noise properties and sky coverage of the LATs for the lensing reconstruction, employing the extended LAT mask after additional edge cuts. For the de-lensing study, we used mock observations with the noise properties of the SAT configuration, while the lensing potential used for de-lensing was estimated from the LAT configuration.
    
\section{Results}   \label{sec:results}
    
\subsection{Harmonic ILC products} \label{sec:hilc_products}
    The Harmonic ILC method outputs a single minimum-variance combination of observations from all frequency channels that have unit responses to the CMB signal by definition. To combine the maps from different frequency channels in spherical harmonic space, the maps were deconvolved to a common resolution, usually to the highest resolution available. However, we used the common resolution of 2.5 arcminutes to avoid any artefacts arising from de-convolving in harmonic space, especially for the 20, 30, and 270 GHz channels as their input resolutions show a huge difference compared to the highest resolution. As mentioned in Sect.~\ref{sec:coverage_noise}, for the maps corresponding to the LAT configuration we used a 5\% masking in the Galactic plane to avoid the most heavily contaminated region. We considered the HILC weights to be the same in the unmasked part of the sky. Dividing the sky into different regions to obtain the spatially varying HILC weights can further improve the component separation. 
    
    For the LAT configuration, the E modes are signal-dominated up to multipoles of $L=3000$, while the lensing B modes are signal-dominated in the range of $200<L<1500$. The residual foreground only dominates the B modes beyond $L=3000$. In our lensing reconstruction analysis, we therefore used modes up to $L=3000$. To suppress foreground contamination at high $L$, we fitted a simple noise model given by Eq.~(\ref{eq:lowL_noise}) to the residual foreground power spectra of the E and B modes, and we used the fitted theoretical model for inverse weighting of the input maps in Eq.~(\ref{eq:inv_weight}). A detailed description of HILC products and the HILC weights for LAT configuration is provided in Appendix \ref{app:hilc}.
     
    The tensor-to-scalar ratio constraints were derived from the deep observations of the SATs. The SAT maps at each frequency were weighted by a normalised hit-count map. In Fig. \ref{fig:hilc_sat_ps}, we compare the average level of foreground contamination in the 95 GHz and 145 GHz channels, which is approximately four orders of magnitude higher than the residual foreground level in the component-separated maps. The HILC-cleaned B-mode maps and residual foreground for all three foreground models are shown in Fig. \ref{fig:sat_residue}. The foreground residues are about one order of magnitude smaller than the signal.

    \subsection{Mean-field estimation}
\begin{figure}
        \centering
        \includegraphics[width=1.0\linewidth]{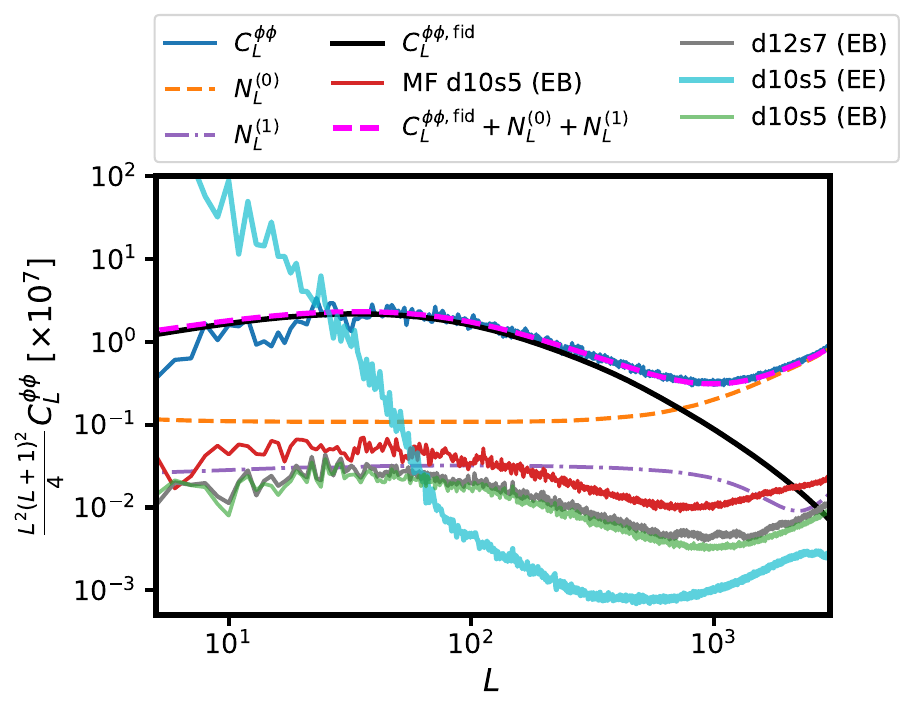} 
        \caption{Angular power spectra of reconstructed lensing field shown as a solid blue line. The mean field effect on the EB estimator for MF maps for d10s5 and HILC-cleaned maps for d10s5 and d12s7 are shown as solid red, green, and grey lines, respectively. The mean-field effect on the EE estimator for HILC-cleaned maps of d10s5 is shown as a solid cyan line. Theoretical lensing power spectra are shown as solid black lines. The dashed magenta line is the sum of theory power spectra and lensing reconstruction noise terms, $N^{(0)}$ and $N^{(1)}$, shown as dashed (orange) and dash-dotted (purple) lines, respectively.}
        \label{fig:mean-field}
    \end{figure}

    We implemented the EB estimator on the HILC-cleaned E-mode and B-mode maps for the LAT configuration and sky coverage. We computed the mean-field effect on lensing reconstruction using a set of 100 simulations as described in Sect. \ref{sec:simulation}. We considered homogeneous instrumental noise and an azimuthally symmetric beam, so the dominant contribution to the mean-field effect arises from masking due to incomplete sky coverage. In addition, the Galactic foreground can be regarded as an anisotropic contamination to the CMB signal, which can further enhance the mean-field bias. To quantify the impact of these foregrounds, we compared two cases. In the first case, we used co-added, foreground-contaminated, noise-free CMB maps of $95+145~\mathrm{GHz}$ before component separation. These co-added maps include the average foreground contribution of the 95 GHz and 145 GHz channels. We then added residual HILC noise realisations to these maps and performed lensing reconstruction. We refer to these foreground contaminated mock observations as the medium-frequency (MF) maps in this paper. The resulting power spectrum of the mean-field effect for MF maps before component separation is shown in red in Fig. \ref{fig:mean-field} for the d10s5 model. In the second case, we evaluated the mean-field effect after component separation for both the d10s5 and d12s7 foreground models. The comparison demonstrates that component separation reduces the overall mean-field contribution by a factor of two. However, the mean-field level is comparable to the $N^{(1)}_L$ reconstruction noise at low multipoles, and it is non-negligible even after component separation. The mean-field contamination does not depend on the foreground models because power spectra for the two models overlap with each other. This is expected as the mean-field effect is mostly dominated by the anisotropies introduced due to masking. The power spectrum of the mean field in all cases is comparable to the $N_L^{(1)}$ bias at low $\ell \lesssim 100$, underlining the necessity of carefully modeling and subtracting it to achieve an unbiased lensing reconstruction. In Fig. \ref{fig:mean-field}, we show the mean-field effect for the EE quadratic estimator of component-separated E-mode maps corresponding to d10s5 model. We can notice that the effect dominates the lensing signal at low $\ell$. One of the main reasons why we did not consider a minimum-variance (MV) combination of all quadratic estimators is discussed in Sect. \ref{sec:sec2.2}. The mean-field effect contributions from same-pair estimators such as the $TT$ and $EE$ estimators makes the reconstruction with an MV estimator noisy at low multipoles, and a careful subtraction is needed in that case.

    \subsection{Foreground bias} \label{sec:foreground_bias}  
    Contamination from Galactic foreground increases the reconstruction noise in the lensing power spectrum coming from Gaussian contribution and also introduces a systematic bias to the reconstructed lensing power spectrum, denoted as $F^{\mathrm{syst}}_L$, which arises from non-Gaussian small scales. We performed lensing reconstruction both before and after component separation to understand these biases in detail. 
    
\begin{figure}
        \centering
        \includegraphics[width=0.9\linewidth]{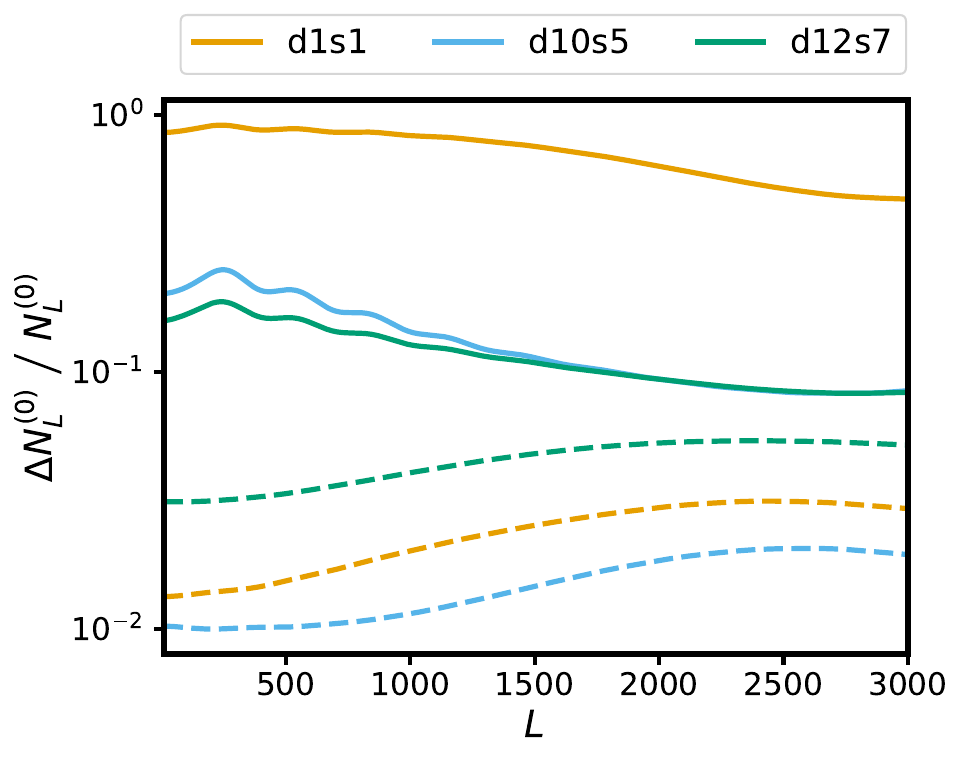}
        \caption{Relative bias in semi-analytic $N_{L}^{(0)}$ noise due to foreground contamination in MF channels (solid lines) before component separation and residual foreground in HILC maps (dashed lines) after component separation for different models.}
        \label{fig:rel_n0_bias}
    \end{figure}

    We defined the fractional change in the zeroth-order reconstruction noise as
    \begin{equation}
        \frac{\Delta N_L^{(0)}}{N^{(0)}_L} = \frac{N^{(0), \mathrm{FG}}_L - N_L^{(0)}}{N_L^{(0)}},
    \end{equation}
    where, $N_L^{(0),\mathrm{FG}}$ is a noise power spectrum with foreground (or residual foreground) contamination, and $N_L^{(0)}$ is a noise spectrum without the presence of foreground contamination. The $N_L^{(0)}$ was computed analytically using Eq. (\ref{eq:nl0_anal_bias}), and $N_L^{(0), FG}$ was computed following Eq. (\ref{eq:n0_anal_foreground}) using foreground (or residual foreground) power spectra. In Fig. \ref{fig:rel_n0_bias}, we compare the biases due to the foreground for MF maps before component separation and for the HILC products after component separation, which contains a residual foreground. We used the same noise realisations in both cases, before and after component separation, while generating simulated maps to only compare the impact of foreground.

    The d1s1 model induces the largest increase in the reconstruction noise prior to component separation, with unity-order changes at low multipoles. For the more realistic d10s5 and d12s7 models with non-Gaussian small scales, the impact on reconstruction noise is less severe, but still non-negligible, leading to a 10-20\% increase in the reconstruction noise across multipoles. In the case of the HILC-cleaned CMB maps, the bias due to residual foreground significantly drops to 2-3\%. Although unmitigated foreground can substantially bias the reconstruction noise, component separation methods are effective in suppressing these contributions. 
  
\begin{figure}
        \centering
        \includegraphics[width=1.0\linewidth]{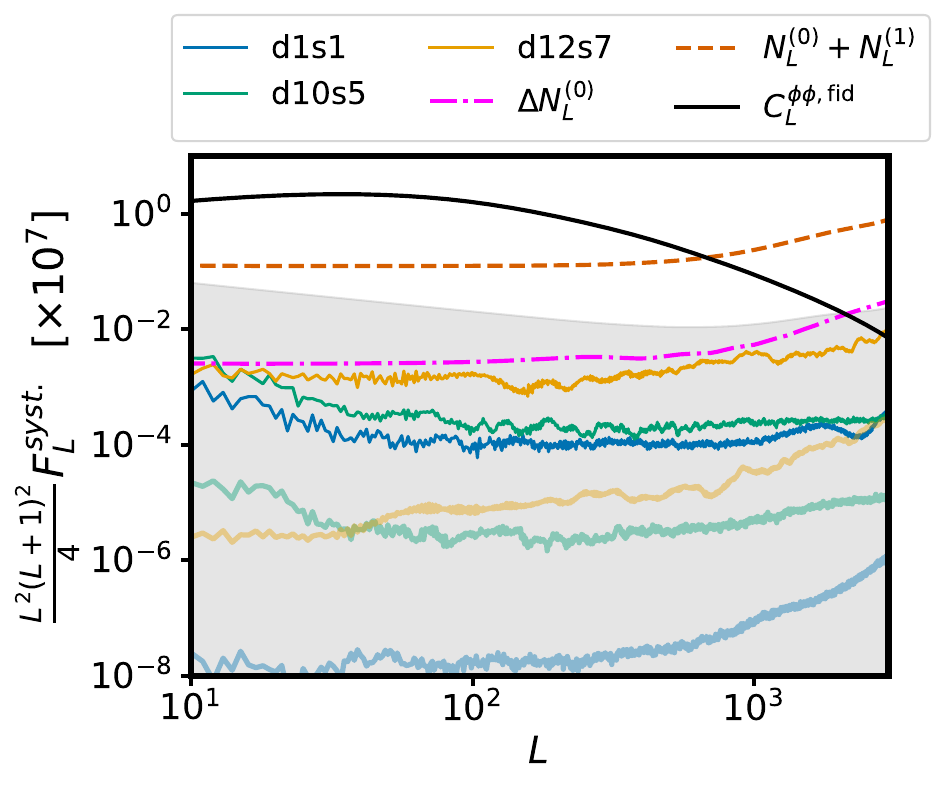}
        \caption{The $F^{syst}_L$ bias is shown for MF maps (dark colours) and HILC-cleaned CMB maps (light colours) for three different models. The black curve shows the fiducial lensing power spectrum, and the dashed red curve is the sum of reconstruction noises, $N^{(0)}+N^{(1)}$. The dash-dotted magenta line is the bias in reconstruction noise, $\Delta N_L^{(0)}$, due to residual Galactic foreground after component separation for the d12s7 model. The shaded grey region represents the cosmic variance limit on the reconstructed lensing power spectrum.}
        \label{fig:flsyst_bias}
    \end{figure}

    In Fig. \ref{fig:flsyst_bias}, we show the systematic bias, $F^{\mathrm{syst}}_L$, for the MF and HILC-cleaned CMB maps. The results show that the systematic bias for the MF map before component separation is below the cosmic variance limit of the total reconstructed lensing power spectrum, including the lensing potential spectrum ($C_L^{\phi\phi}$) and reconstruction noise ($N_L^{(0)}+N_L^{(1)}$). The component separation further reduces the systematic bias to three orders of magnitude lower than $N_L^{(0)}$, making it negligible for our analysis. For highly non-Gaussian models such as the d12s7 model, we see the largest $F_L^{syst}$ bias. A comparison of Fig. \ref{fig:rel_n0_bias} and Fig. \ref{fig:flsyst_bias} shows that the $\Delta N_L^{(0)}$ bias in reconstruction noise is more pronounced for foreground models dominated by the small-scale Gaussian structure, while models with stronger non-Gaussianity contribute significantly to the systematic bias in the reconstructed power spectra. In Fig. \ref{fig:flsyst_bias}, we also present the bias coming from the Gaussian contribution of $\Delta N_L^{(0)}$ for the d12s7 model after component separation, which is almost two orders of magnitude larger than corresponding systematic bias arising from non-Gaussian foreground residues, and it is comparable to the cosmic-variance limit.

\subsection{De-lensing and $r$ constraint}
    
    Implementing Eq. (\ref{eq:lensing_Btemp}), we obtained lensing B-mode templates using Wiener-filtered, foreground-cleaned (HILC) E-mode maps and the reconstructed lensing potential field over the LAT configuration sky coverage ($f_{sky}=0.37$). We included residual foreground and noise power spectra in the Wiener filtering. We used this template to de-lens the B-mode map using Eq. (\ref{eq:delens_Bmode}) for the deep-field observations from the SAT configuration. Subtracting the B-mode template does not fully remove the lensing-induced modes, leaving some residual lensing signal as shown in Fig. \ref{fig:delensed_bmode}. The de-lensed B-mode maps also include noise and residual foreground components, which contribute to the measured power spectra. In Fig. \ref{fig:hilc_sat_ps}, we show the average power spectrum of de-lensed CMB maps for a set of 100 simulations. The delensed power-spectra level is one order of magnitude larger than the residual noise and foreground-spectra level, which shows that the de-lensed maps are dominated by residual lensing signal.

    To estimate the tensor-to-scalar ratio, we performed a likelihood analysis of the de-lensed B-mode power spectrum using Eq. (\ref{eq:likelihood}). The model power spectrum is given in Eq. (\ref{eq:model_cl}). The full covariance matrix was estimated from 400 purely Gaussian realisations of the foreground, allowing us to propagate the impact of residual foreground into the likelihood. The residual foreground power spectra, $F_\ell^{res}$, in Eq. (\ref{eq:model_cl}) was obtained by taking an average over 400 purely Gaussian realisations of Galactic-foreground maps. The steps followed to simulate the Gaussian foreground are explained in Sect. \ref{sec:galactic_foreground}. To show the off-diagonal contributions of the covariance matrix, we present the corresponding correlation matrices for our three models in Fig. \ref{fig:likelihood_corr_mat}. The model d10s5 shows the largest off-diagonal contribution in the correlation matrix.
    
\begin{figure}
        \centering
        \includegraphics[width=1.\linewidth]{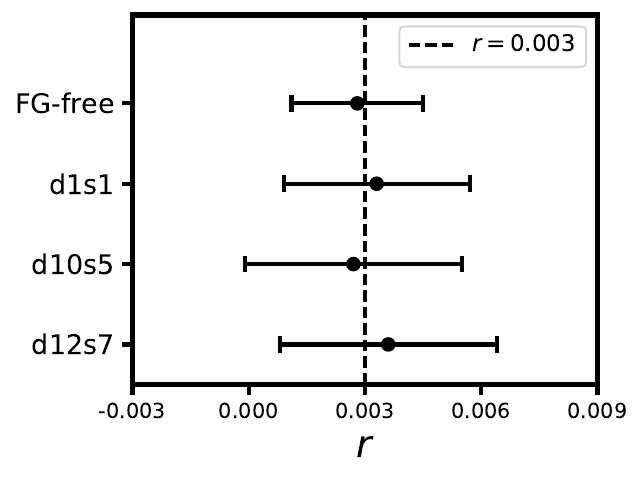}
        \caption{Mean value of the tensor-to-scalar ratio, $r$, and its standard deviation, $\sigma(r)$, estimated from 100 simulations for the three considered foreground models and FG-free maps. The vertical dashed line corresponds to the fiducial value of $r$.}
        \label{fig:mean_r_estimate}
    \end{figure}

    We implemented the full pipeline on four sets of simulated maps: one set with foreground-free (FG-free) maps (see Sect. \ref{sec:galactic_foreground}) and three sets each including foreground contamination based on the three models. All the maps in each set were passed through the HILC pipeline, lensing reconstruction, template building, and de-lensing pipeline before we performed the likelihood analysis. We present the mean estimate of the tensor-to-scalar ratio, $r$, obtained from a set of 100 simulations each for all four cases. The standard deviation of the mean shown in Fig. \ref{fig:mean_r_estimate} represents the spread of the mean of the posterior of each simulation. The error on the mean, $r,$ varies slightly between the models, with d12s7 showing the largest uncertainty due to its higher level of complexity. No significant deviation from the fiducial value is observed, demonstrating that the residual foreground biases in lensing reconstruction after component separation and correcting for the biases in the analysis do not have a significant impact on the estimation of $r$. The presence of residual foreground in SAT maps increases the uncertainty of the $r$ measurement compared to the foreground-free case.

\subsection{Error budget on $r$}
    The additional B-mode components from residual instrumental noise and residual foreground contributes to the uncertainties in parameter estimation. To estimate contributions from different components to the error on $r$ estimation, we used a Fisher matrix formalism. We considered the full-sky log-likelihood scaled by a factor of $f_{sky}$  for the cut-sky power spectrum estimator with a given model power spectrum of Eq. (\ref{eq:model_cl}). 
    Then, we obtained the uncertainty on the $r$ parameter from Fisher information matrix,
    \begin{align}
        \begin{split}\label{eq:sigma_r}
            \sigma^2(r) & = \left< -~\frac{\partial^2 \ln \mathcal{L}}{\partial r^2} \right> ^{-1} \\
                    \quad = & \frac{1}{f_{sky}} \sum_{\ell_{b}} \frac{2}{(2\ell_b + 1)\Delta\ell} \left( \frac{C_{\ell_b}^{tens, r=1}}{r\,C_{\ell_b}^{tens,r=1} + C_{\ell_b}^{res} + N^{res}_{\ell_b} + F^{res}_{\ell_b}} \right).
        \end{split}
    \end{align}
    Here, we used binned power spectra with a bin width of $\Delta\ell$. The summation is done over the multipoles at the centre, $\ell_b$, of each bin. We considered the bins within $50\leq \ell_b \leq 300$ with a width of $\Delta\ell=30$.
    
\begin{figure}
        \centering
        \includegraphics[width=1.\linewidth]{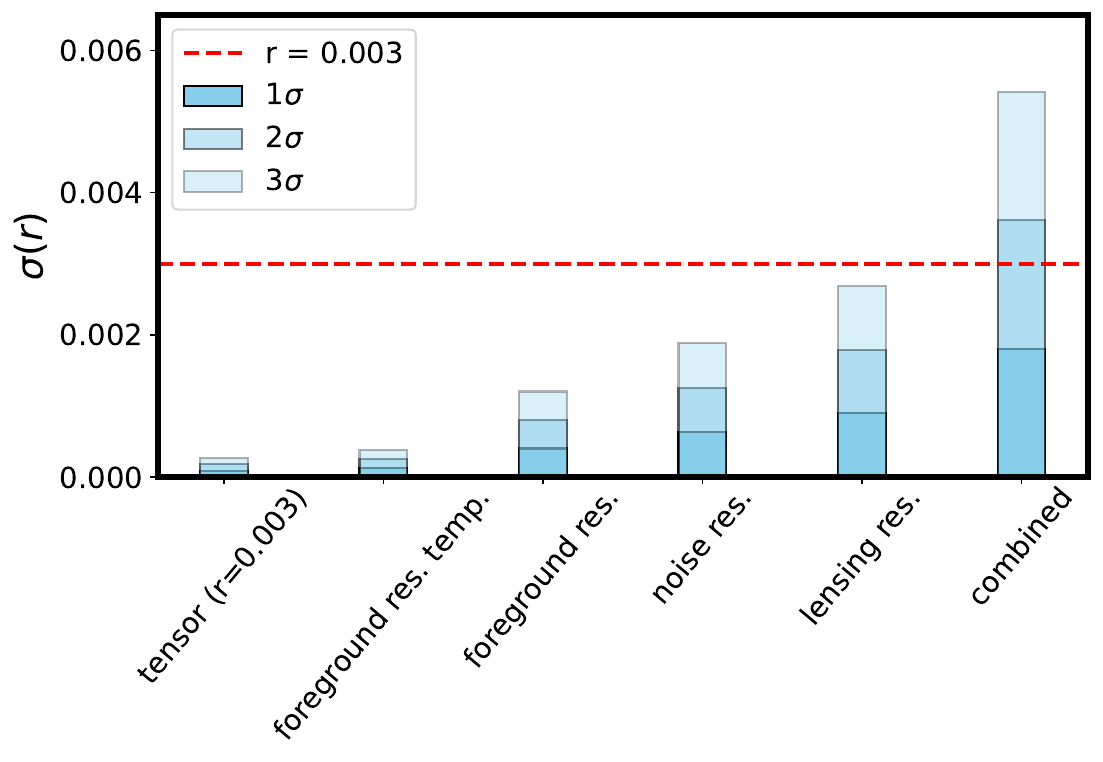}
        \caption{Contribution to $\sigma(r)$ from different components of observed B modes in the presence of foreground (d10s5).}
        \label{fig:sigma_r}
    \end{figure}

    In Fig.~\ref{fig:sigma_r}, we show the individual contributions to the total uncertainty in the tensor-to-scalar ratio, $r$, arising from different components present in the de-lensed $B$-mode power spectrum. The total uncertainty, $\sigma(r)$, represents the expected statistical error due to the cosmic-variance limit on the measured de-lensed power spectrum for a single realisation. We obtained the uncertainty for each component of the model power spectrum in Eq. (\ref{eq:sigma_r}) by considering contribution from that particular component while setting all other components to zero. This approach allowed us to isolate the effect of each source of uncertainty.

    The tensor ($r=0.003$) term in Fig. \ref{fig:sigma_r} reflects the fundamental limit imposed by the primordial $B$-mode signal itself, corresponding to the cosmic variance of the tensor component for a fiducial value of $r=0.003$ considered in this paper. To quantify the impact of residual foreground in the template map used for de-lensing, we computed the difference in the variance of $r$, obtained using component-separated CMB maps with residual foregrounds and using analogous maps with subtracted residual foreground (foreground-free HILC CMB maps). This uncertainty level from foreground residue in template building --shown in Fig. \ref{fig:sigma_r} as foreground res. temp.-- gives a quantitative measure of how residual foreground contamination in the B-mode template propagates into the constraint on $r$. The foreground res. and the noise res. terms arise from residual foreground and instrumental noise, respectively, which is present in the component-separated B-mode SAT maps. The lensing res. term captures the remaining $B$-mode power due the incomplete removal of lensing-induced $B$-modes. Finally, the combined uncertainty shows the total uncertainty when all contributions are included simultaneously, illustrating how each component cumulatively increases $\sigma(r)$. Note that the combined uncertainty is not simply the sum of each individual contribution from different components.

\section{Conclusions}     \label{sec:conculsion}
    In this work, we studied the impact of Galactic foreground and its residuals on CMB gravitational-lensing reconstruction, the de-lensing of CMB maps, and primordial gravitational-wave searches. Using simulations of Galactic foreground based on three different models, we quantified the mean-field effect, the bias in the reconstruction noise, and the non-Gaussian systematic bias. We demonstrated that the mean-field bias arises primarily from masking anisotropies, and that its amplitude for the EB quadratic estimator is comparable to the $N^{(1)}_L$ bias. We further showed that foreground affects the reconstruction of a lensing power spectrum in two different ways. The Gaussian component of the foreground primarily increases the reconstruction noise at intermediate and high multipoles, while non-Gaussian small-scale features induce a systematic bias at low multipoles. 
    
    We demonstrated that component separation is an essential step for both lensing power-spectrum reconstruction and tensor-to-scalar ratio estimation. In lensing reconstruction, component separation reduces the bias in $N_L^{(0)}$ reconstruction noise, as well as systematic biases from non-Gaussian small-scale features. As shown in Fig.~\ref{fig:hilc_sat_ps}, for the estimation of the tensor-to-scalar ratio, the de-lensed spectra are initially dominated by foreground contamination, where the cleanest frequency channels before component separation exhibit strong foreground emission, which is reduced by nearly two orders of magnitude after component separation. We see that the increase in reconstruction noise, $\Delta N^{(0)}_L$, for the most complex model, d12s7, is comparable to a cosmic-variance limit at high multipoles. Therefore, we included the bias in the Wiener filtering when generating the template B-mode map in order to obtain optimal de-lensing efficiency and improved constraints on the tensor-to-scalar ratio. For the three considered foreground models, the estimated values of $r$ from the likelihood analysis are consistent with the fiducial input value of $r = 0.003$. No significant bias in the mean estimated value of $r$ is observed due to residual Galactic foreground. We observe only an $\sim10\%$ deviation in the case of the highest complexity model, d12s7, which could be due to the fact that the covariance matrix computed from purely Gaussian foreground does not fully capture the contribution of residual foreground in the delensed spectra. 
    
    The average uncertainty on the tensor-to-scalar ratio across the three models is about $70-80\%$ of the fiducial value in our analysis, with the d12s7 model showing the largest uncertainty. We compare our results with those from \citet{Beck_2020} and find approximately twice the level of uncertainty on the value of $r$. This difference can be attributed to the fact that we used the full Hamimeche and Lewis likelihood without the Gaussian approximation, thereby including off-diagonal correlations that tend to broaden the $\sigma(r)$. In addition, we considered more realistic and updated models of the Galactic foreground, such as d10s5 and d12s7, which include higher levels of complexity and non-Gaussian structures on small angular scales. Furthermore, we employed a non-parametric component-separation technique, the Harmonic ILC method, which allowed us to avoid relying on prior assumptions about foreground.

    In a similar work, \citet{Bianchini_2025} investigated the impact of Galactic-foreground complexity on the estimation of the tensor-to-scalar ratio by comparing multiple component-separation pipelines applied to simulated CMB-S4--like data. Their analysis focused on biases and uncertainties in $r$ arising from foreground cleaning and pipeline choices. In contrast, our work focused on the impact of non-Gaussian Galactic-foreground residues on CMB lensing reconstruction and the resulting constraints on $r$ using a gradient-order, template-based de-lensing method. \citet{Bianchini_2025} employed an iterative de-lensing procedure that removes nearly $90\%$ of the lensing signal, yielding tighter constraints on $r$. Another key difference in the methodologies is that, rather than marginalising over residual foregrounds as in \citet{Bianchini_2025}, we included the residual foreground contribution directly in the covariance matrix and in the model power spectra. Together, these studies provide complementary perspectives on how foreground complexity propagates through different stages of CMB polarisation analyses.

    As we show in Fig.~\ref{fig:sigma_r}, a $3\sigma$ detection of primordial gravitational waves with $r \sim 10^{-3}$ using a simple quadratic $EB$ estimator in the de-lensing of the B-mode map is not achievable for CMB-S4--like experiments. The dominant contribution to the total uncertainty arises from the lensing residuals present in the de-lensed $B$-modes. Using a polarisation-based $EB$ estimator, we were able to remove approximately $65\%$ of the lensing-induced $B$-mode power. Finding optimal methods for de-lensing was beyond the of scope of our analysis; however, there are approaches enabling the improvement of the de-lensing efficiency. One such approach, especially relevant for small angular scales, is to combine the internally reconstructed lensing potential with other large-scale structure (LSS) tracers to generate the B-mode template used for de-lensing \citep{Manzotti_2018, Litebird_2019, bicep_spt_2021, Hertig_2024}. One can also use iterative de-lensing methods \citep{Carron_MAP_2017, Belkner_2024}, removing up to about $90\%$ of the lensing contamination. Such improvements would significantly reduce the contribution of lensing residuals to $\sigma(r)$, potentially enabling a $3\sigma$ detection of tensor modes down to $r \sim 10^{-3}$. In such cases, residuals of the Galactic foreground, together with instrumental noise, provide a dominant contribution to the errors on the tensor-to-scalar ratio. For this reason, robust component separation and accurate correcting for the foreground residues for CMB lensing reconstruction analysed in our work will be highly relevant for the extraction of unbiased lensing potential and the detection of the primordial B-mode signal.

%-----------------------------------------------------------------

\begin{acknowledgements}
       The authors acknowledge the use of the public software packages : CAMB, healpy, lenspyx, cmblensplus, plancklens, NaMaster, emcee, PySM3, Numpy, Scipy and matplotlib. We thank Radek Stompor, Dominic Beck, Carlo Baccigalupi and Anto I. Lonappan for helpful comments and suggestions on the results of this work. We thank Giuseppe Puglisi and Tuhin Ghosh for their suggestions on Galactic foreground modeling. We also thank Jacques Delabrouille, Julien Carron, Shamik Ghosh and Chandra Shekhar Saraf for their useful comments on  delensing methods and on likelihood analysis techniques. KD acknowledges partial support from STER programme BPI/STE/2021/1/00033/U/00001 funded by NAWA. The authors are thankful for the computational resources provided by the CI\'S NCBJ cluster. This work used the NASA Astrophysics Data System Bibliographic Services.
\end{acknowledgements}

\bibliographystyle{aa}
\bibliography{bibfile}

\begin{appendix}
    \section{HILC method implementation} \label{app:hilc}
    Assuming multi-frequency foreground contaminated CMB observations in the $N_c$ frequency bands and frequency maps convolved with an instrumental beam function, the observed signal for the $i^{th}$ frequency band, where $i \in 1,2,\ldots,N_c$, in terms of harmonic coefficients of a given mode-index $(\ell,m)$, can be written as
    \begin{equation}
        a^{f,i}_{\ell m} = B^i_\ell \left( a^{c,i}_{\ell m} + a^{s}_{\ell m}\right) + a^{n,i}_{\ell m} 
    ,\end{equation}
    where the coefficients $a^{f,i}_{\ell m}$, $a^{c,i}_{\ell m}$, $a^s_{\ell m}$ and $a^{n,i}_{\ell m}$ denotes the total observed signal, net foreground emission, CMB signal and detector noise contribution, respectively. $B_\ell^i$ is the beam transfer function of the instrument for $i^{th}$ frequency band. Before applying HILC method to these coefficients, all the maps are first deconvolved with a beam of frequency-specific resolution and then convolved with common beam, generally to the highest available resolution amongst observed frequencies. Then, the cleaned CMB signal can be obtained as a linear combination of all $N_c$ observed frequency maps in harmonic space, 
    \begin{equation}
        a^{Clean}_{\ell m} = \sum_{i=1}^{N_c} w^i_\ell \frac{B^o_\ell}{B^i_\ell} a^{f,i}_{\ell m}
    ,\end{equation}
    where $B^o_\ell$ is the common output beam function and $w^i_\ell$ is the corresponding weight function of $i^{th}$ band for a given multipole $\ell$. The weights $w^i_\ell$ are obtained by minimising the variance of the clean CMB signal under the constraint
    \begin{equation}
        \sum_i w_i = 1.  \label{eq:unity_constraint}
    \end{equation}
    The constraint on the weights in Eq. (\ref{eq:unity_constraint}) makes sure that the CMB signal is preserved. The minimum variance weight can be written as \citep{Tegmark_2003, Kim_2009}
    \begin{equation}
        w^i_\ell =  \frac{\sum_{j} [\mathbf{\hat{C}}^{-1}_{\ell}]^{ij}}{\sum_{ij} [\mathbf{\hat{C}}^{-1}_{\ell}]^{ij}}
    .\end{equation}
    Here, $\mathbf{[\hat{C}}^{-1}_{\it \ell}]^{\it ij}$ denotes an element of the inverse of the covariance matrix having the $i^{th}$ row index and $j^{th}$ column index. Corresponding element of the covariance matrix is defined as
    \begin{equation}
        \mathbf{[\hat{C}_{\ell} ]}^{ij} =  \frac{1}{2 \ell+1}\sum_{m=-\ell}^{\ell}a^{f,i}_{\ell m}a^{f,j*}_{\ell m}
    .\end{equation}

    The HILC-cleaned maps still contains HILC noise and residual Galactic foreground signal. The residual noise and residual foreground maps are defined as the HILC weighted sum of noise and Galactic foreground contributions of each frequency channel. We estimate the residual noise and residual foreground and their power spectra as
    \begin{align}
        n^{res}_{\ell m} = \sum_{i} w_\ell^i a^{n,i}_{\ell m}, \hspace*{0.5cm} f^{res}_{\ell m} = \sum_i w^i_\ell \left(\frac{B^o_\ell}{B^i_\ell}\right) a^{f,i}_{\ell m} , \label{eq:residual_map} \\
        N_\ell^{res} = 1/(2\ell+1) \sum_m |n^{res}_{\ell m}|^2 ,   \hspace*{0.5cm}
        F_\ell^{res} = 1/(2\ell+1) \sum_m |f^{res}_{\ell m}|^2 . \label{eq:noise_residue_ps}
    \end{align}
    Here, $n^{res}_{\ell m}$ and $f^{res}_{\ell m}$ denote residual noise and foreground spherical harmonic coefficients, respectively, and $N^{res}_\ell$ and $F^{res}_\ell$ are corresponding estimators of the power spectra.

\begin{figure}
        \centering
        \includegraphics[width=1.\linewidth]{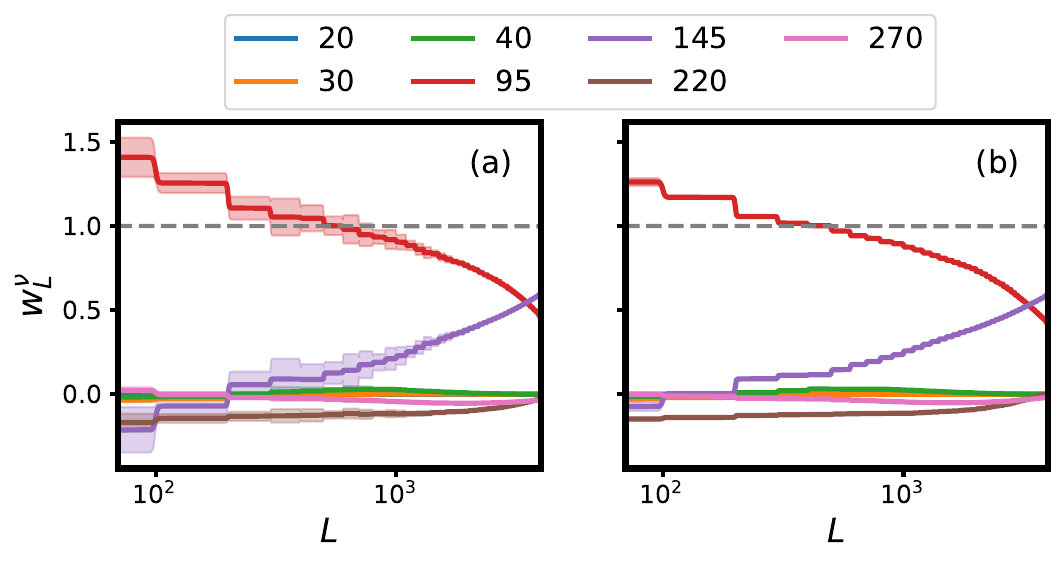}
        \caption{Mean and standard deviation of HILC-derived weights over 100 realisation for d10s5 model in LAT configuration ($f_{sky}=0.37$).}
        \label{fig:hilc_weights}
    \end{figure}

    In Fig. \ref{fig:hilc_weights}, we present the mean HILC weights and their standard deviation for the frequency channels in LAT configuration in the case of d10s5 Galactic foreground model computed from a set of 100 realisations. We see that, the cleanest channels $95$ and $145$ GHz has the highest contributions in the final HILC product. The weights for rest of the channels are closer to zero and do not contribute at high multipoles. The weights are estimated in a band of multipoles which is seen as step-like pattern in the plot.
    
\begin{figure}
        \centering
        \includegraphics[width=1.\linewidth]{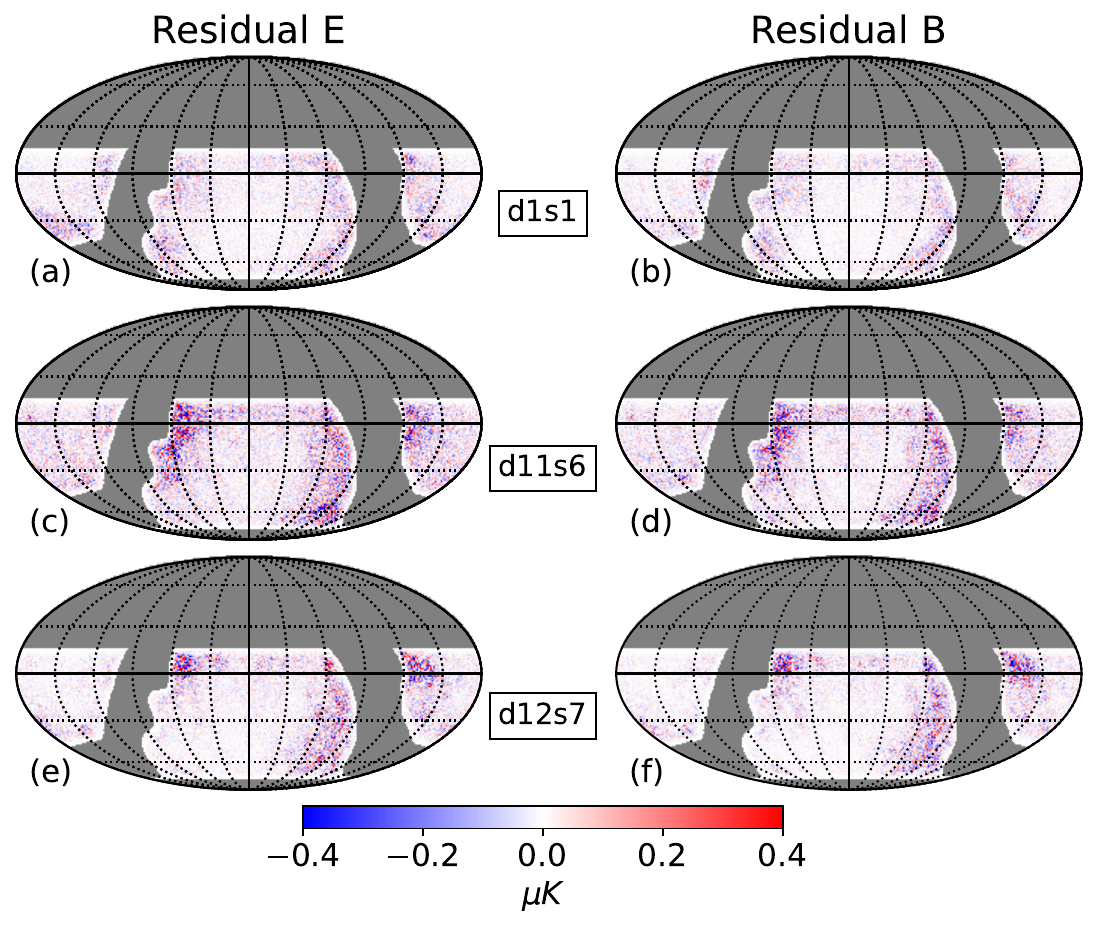}
        \caption{Residual foreground in HILC-cleaned E and B modes maps for one realisation in LAT configuration ($f_{sky}=0.37$).}
        \label{fig:fgres_map_lat}
    \end{figure}

\begin{figure*}
         \centering
        \includegraphics[width=0.75\textwidth]{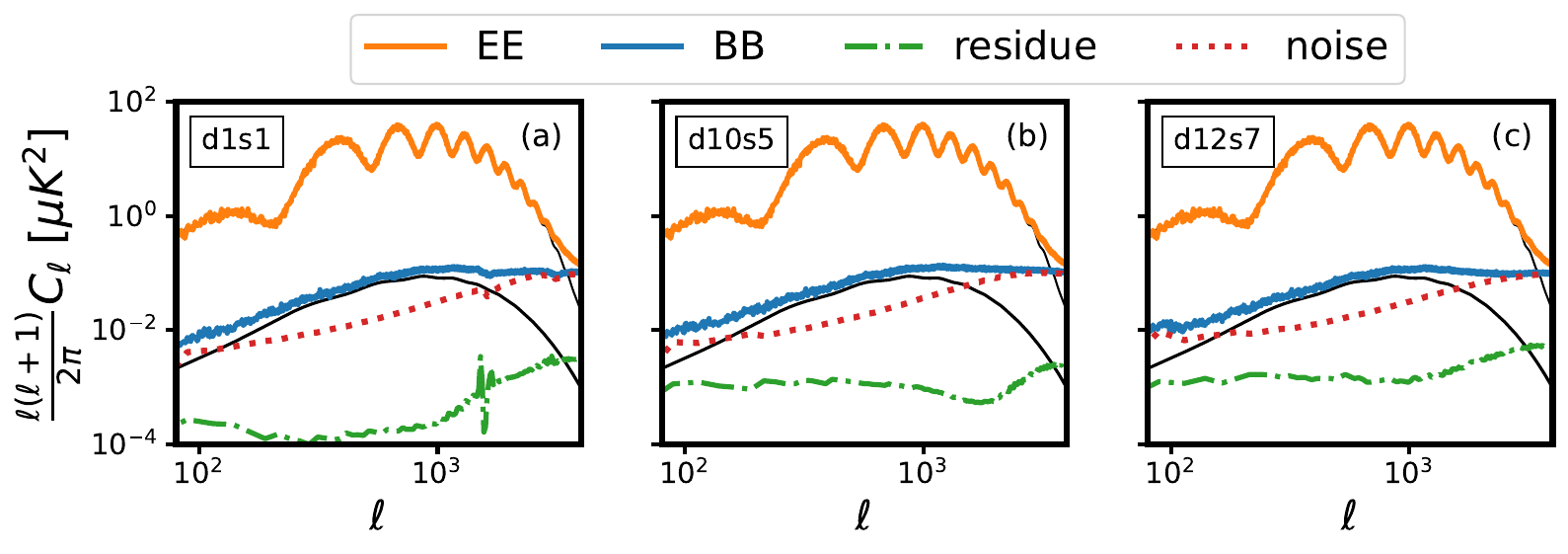}
        \caption{Angular power spectra for CMB E and B-mode maps cleaned using the HILC method for the LAT configuration and sky coverage. The orange and blue solid zigzag lines are average power spectra over 100 simulations for the E and B-mode maps, respectively. The theoretical input power spectra are shown as black solid line. The map residual B-mode noise and foreground spectra for B modes are shown as dotted and dash-dotted lines, respectively.}
        \label{fig:hilc_lat_ps}
    \end{figure*}

    The E- and B-mode HILC-cleaned LAT maps contain residual noise and residual foreground, which can be computed using Eqs. (\ref{eq:residual_map}) and (\ref{eq:noise_residue_ps}). In Fig. \ref{fig:fgres_map_lat} we compare the residual foreground for E and B modes from different foreground models in LAT configuration. Figure~\ref{fig:hilc_lat_ps} shows a comparison of the residual noise and residual Galactic foreground power spectra in the HILC B-mode products with the theoretical E- and B-mode input power spectra. The residual noise and residual foreground in the E-mode power spectra are of the same order of magnitude. 
    
    \begin{figure*}
        \centering
        \includegraphics[width=0.7\textwidth]{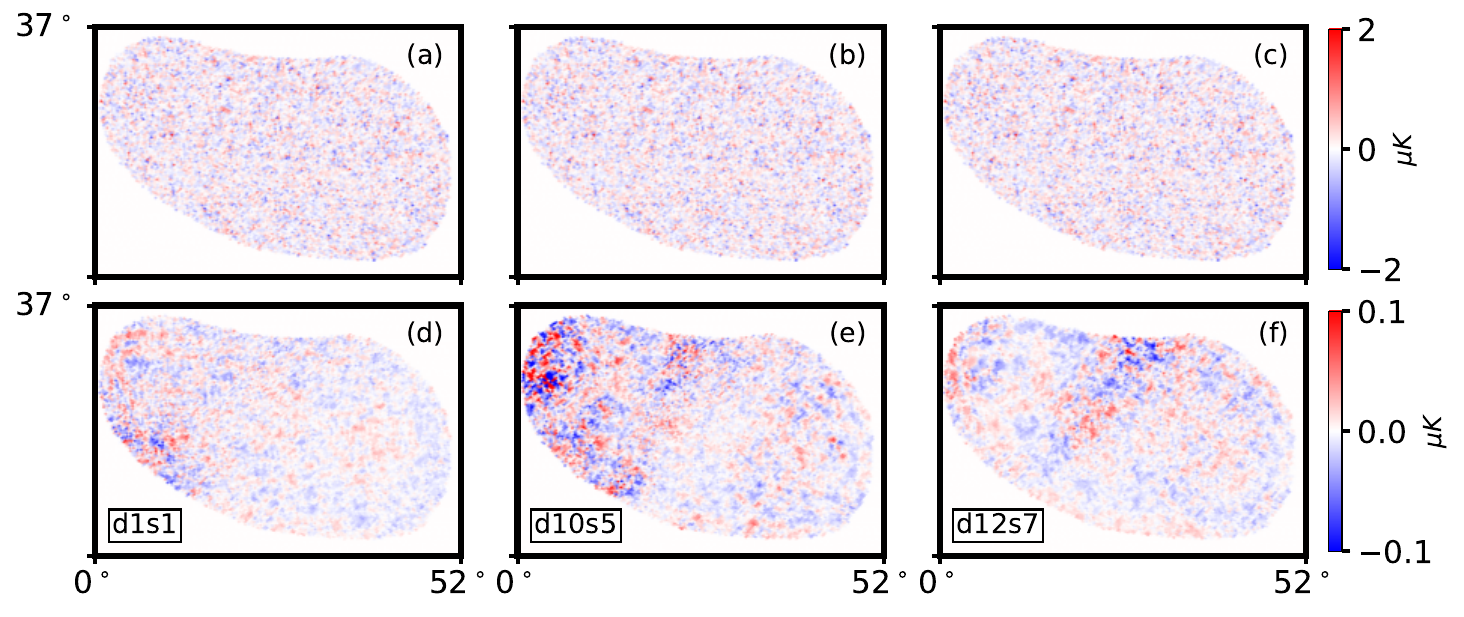}
        \caption{HILC-cleaned CMB maps (upper row) and foreground residual maps (lower row) for SAT configuration and sky coverage for the Galactic foreground models d1s1 (left), (b) d10s5 (middle) and d12s7 (right). These cleaned maps (upper row) includes residual HILC noise and foreground defined by Eq. (\ref{eq:residual_map}).}
        \label{fig:sat_residue}
    \end{figure*}

    \section{Covariance matrix for likelihood analysis}
\begin{figure}
        \centering
        \includegraphics[width=.9\linewidth]{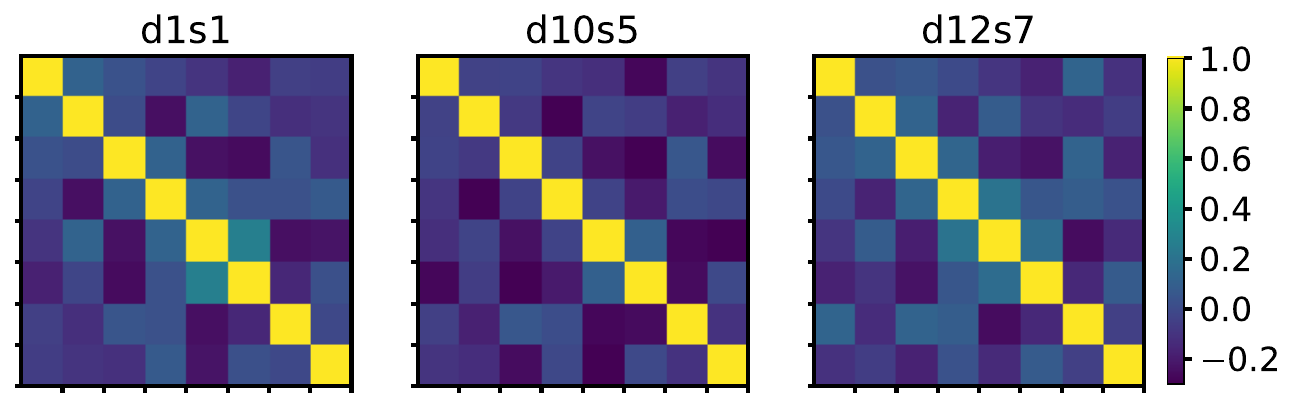}
        \caption{Correlation matrices for the covariance matrix in Eq. (\ref{eq:cov_mat}) for the three foreground models. The full covariance matrices have off-diagonal contributions.}
        \label{fig:likelihood_corr_mat}
    \end{figure}

    The covariance matrices defined by Eq. \ref{eq:cov_mat} captures non-zero off-diagonal contributions for the three foreground models considered. We computed correlation matrix for the covariance matrix shown in Fig. \ref{fig:likelihood_corr_mat} and we see that off-diagonal terms are non-negligible. Therefore, we considered the inverse of full covariance matrix in the likelihood analysis to constraint the tensor-to-scalar ratio.
    
    \section{Delensed B-mode maps}
\begin{figure*}
        \centering
        \includegraphics[width=0.7\linewidth]{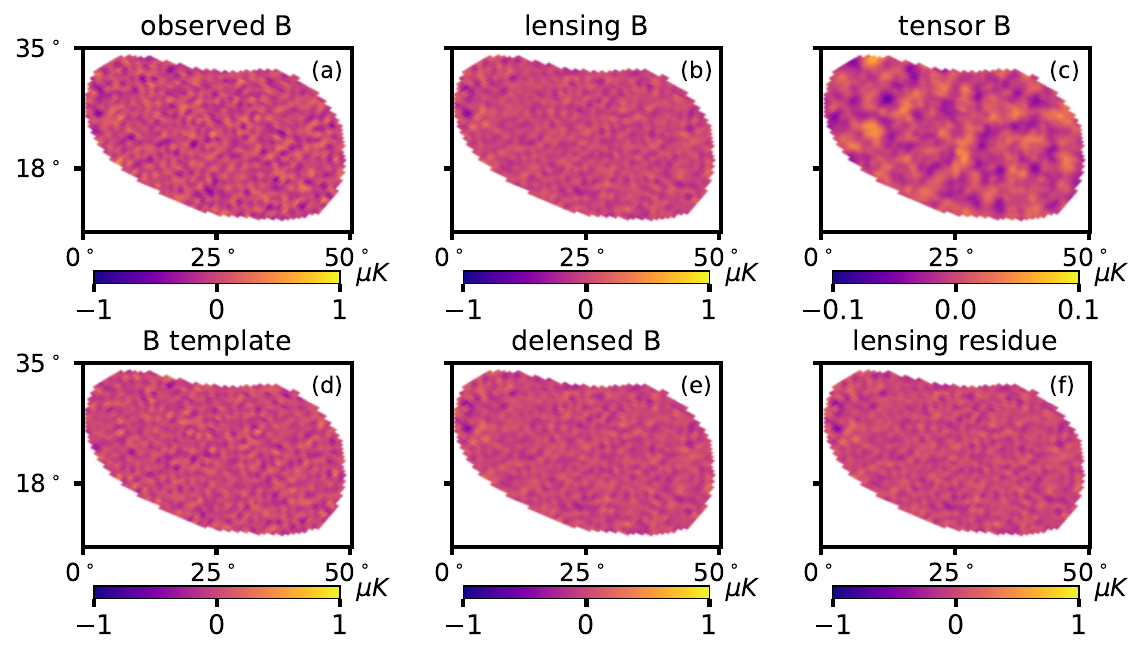}
        \caption{Different components of the CMB B-mode maps for SAT configuration and sky coverage. All the maps only includes multipoles in the range $30 \leq \ell \leq 300$. Observed B-mode map in top-left panel includes residue after component separation noise and foreground for the d10s5 model. The remaining maps are noise-free and foreground-free.}
        \label{fig:delensed_bmode}
    \end{figure*}

    Delensed B-mode maps includes different components of B-mode signal as mentioned in the power spectrum model in Eq. \ref{eq:model_cl}. In Fig. \ref{fig:delensed_bmode}, we show each different component separately to compare their level of contribution to the delensed B-mode map.
\end{appendix}

\end{document}